\documentclass[prd,aps,nofootinbib,notitlepage
]{revtex4}
\usepackage{graphicx,epsf,amsfonts,amssymb,amsbsy}
\newcommand{\mat}{\left ( \begin{array}}
\newcommand{\emat}{\end{array} \right )}
\newcommand{\vect}{\left ( \begin{array}{c}}
\newcommand{\evect}{\end{array} \right )}

\begin{document}

\title{ 
Detailed study of the dual symmetry of dense quark matter within the effective action formalism }
\author{T. G. Khunjua $^{1)}$, K. G. Klimenko $^{2)}$, and R. N. Zhokhov $^{3)}$ }
\vspace{1cm}

\affiliation{$^{1}$ The University of Georgia, GE-0171 Tbilisi, Georgia}
\affiliation{$^{2)}$ State Research Center
	of Russian Federation -- Institute for High Energy Physics,
	NRC "Kurchatov Institute", 142281, Protvino, Moscow Region, Russia}
\affiliation{$^{3)}$  Pushkov Institute of Terrestrial Magnetism, Ionosphere and Radiowave 
Propagation (IZMIRAN), 108840 Troitsk, Moscow, Russia}


\begin{abstract}
The symmetry properties of the phase diagram of dense quark matter composed of 
$u$ and $d$ quarks with three colors have been investigated in the framework of massless 
(3+1)-dimensional Nambu--Jona-Lasinio (NJL) and QCD models. It turns out that in the 
presence of baryon $\mu_B$, isospin $\mu_I$, chiral $\mu_5$ and chiral isospin $\mu_{I5}$ 
chemical potentials, the Lagrangians of these models are invariant under the so-called dual 
transformation ${\cal D}$. In this paper, it has been shown that the path integration measure of the corresponding 
partition functions is also invariant under ${\cal D}$. 

Consequently, the entire NJL model (or QCD) thermodynamic potentials are dually symmetric. In particular, it means that in the 
total $(\mu_B,\mu_I,\mu_5,\mu_{I5})$-phase portraits of these models the chiral symmetry 
breaking (CSB) and charged pion condensation (PC) phases are arranged dually conjugate 
(or symmetrically) to each other. Dual symmetry between CSB and charged PC
phases of quark matter are a fundamental property of massless dense QCD; it should be observed
in the framework of any approximation to the QCD phase diagram, and have important implications for studies of the phase structure. 
\end{abstract}

\maketitle

\section{Introduction}

Dense quark matter described by Nambu--Jona-Lasinio (NJL) models with isospin and chiral imbalances exhibits a discrete duality $\mathcal{D}$ between chiral symmetry breaking (CSB) and charged pion condensation (PC) phenomena: under a Pauli--Gursey--type rotation of quark fields and a permutation of chemical potentials $\mu_I \leftrightarrow \mu_{I5}$, the two phases are mapped onto one another. This duality has been established in a series of works on $(1+1)$- and $(2+1)$-dimensional four-fermion models \cite{thies_a,thies_b,Thies2_a,Thies2_b,Thies2_c,ebert,cao,ekkz, kkz, kkz17}, on the realistic $(3+1)$-dimensional NJL model with two flavors and three colors \cite{kkz18,kkz18-2,kkz19,kkz}, and extended to the two-color case with diquark interactions \cite{kkz20_a,kkz20_b}; further dual structures of a three-color NJL model with color superconductivity were investigated in \cite{u,u2}. Most recently, the duality was shown to hold at the level of the two- and three-color QCD Lagrangian itself \cite{Khunjua:2024kdc_a,Khunjua:2024kdc_b}.

While the invariance of the Lagrangian was proven in those works, the invariance of the thermodynamic potential (TDP) was only implied. In the present paper we close this gap. Within the effective-action formalism we prove that the TDP of massless QCD and QCD-like models is invariant under $\mathcal{D}$, and we further show that the path-integral measure in the generating functional is invariant under $\mathcal{D}$, so anomalies do not destroy the duality. Together these results imply that the duality between CSB and charged PC is an intrinsic property of QCD and its effective models, holding beyond the mean-field or leading-$N_c$ approximations.

The paper is organised as follows. Section~\ref{a} sets up the NJL model with the chemical potentials $\mu_B, \mu_I, \mu_{I5}, \mu_5$ and recalls the mean-field duality. Section~\ref{aaa3} formalises the dual transformation in the effective-action framework. Section~\ref{aaa3a} extends the analysis to the NJL model with diquark channels, and Sec.~V to two-flavor QCD itself. Appendices~\ref{ApA}--\ref{ApD} collect technical material on the symmetries of the effective action and on the dual invariance of the path-integral measure.

\section{Duality between CSB and charged PC in the mean-field approximation} \label{a}

We work with the three-color ($N_c=3$), two-flavor, massless NJL Lagrangian extended by baryon, isospin, chiral-isospin, and chiral chemical potentials (cf.~Refs.~\cite{kkz18-2,Khunjua:2024kdc_a,Khunjua:2024kdc_b}):
\begin{equation}
L_{NJL}=i\overline{\psi} \gamma^{\mu}\partial_{\mu} \psi+\overline{\psi}\gamma^0{\cal M}\psi+
\frac{G}{N_c}\Big\{(\overline{\psi}\psi)^2+ (i\overline{\psi}\gamma^5\vec\tau\psi)^2\Big\},\label{1}
\end{equation}
with the chemical-potential matrix
\begin{eqnarray}
{\cal M}=\frac{\mu_B}{3}+\frac{\mu_I}2\tau_3+\frac{\mu_{I5}}2\gamma^5\tau_3+\mu_5\gamma^5,
  \label{200}
\end{eqnarray}
where $\psi$ is a flavor doublet and a color triplet (in Sec.~\ref{aaa3a} a color-doublet variant is treated separately), and $\tau_k$ ($k=1,2,3$) are Pauli matrices acting in flavor space.

Lagrangian~(\ref{1}) is invariant under the continuous group $U_B(1)\times U_{I_3}(1)\times U_{AI_3}(1)$, whose generators read \footnote{\label{f1,1}
For the following, recall that~~
$\exp (\mathrm{i}\alpha\tau_3)=\cos\alpha
+\mathrm{i}\tau_3\sin\alpha$,~~~~
$\exp (\mathrm{i}\alpha\gamma^5\tau_3)=\cos\alpha
+\mathrm{i}\gamma^5\tau_3\sin\alpha$.}
\begin{eqnarray}
U_B(1):~\psi\to\exp (\mathrm{i}\alpha/3)\psi;~
U_{I_3}(1):~\psi\to\exp (\mathrm{i}\alpha\tau_3/2)\psi;~
U_{AI_3}(1):~\psi\to\exp (\mathrm{i}\alpha\gamma^5\tau_3/2)\psi.
\label{2}
\end{eqnarray}
The corresponding conserved densities are $n_B=\langle\overline\psi\gamma^0\psi\rangle/3$, $n_I=\langle\overline\psi\gamma^0\tau_3\psi\rangle/2$, and $n_{I5}=\langle\overline\psi\gamma^0\gamma^5\tau_3\psi\rangle/2$. The chiral chemical potential $\mu_5$, by contrast, does not correspond to a conserved charge; it parameterises matter with a fixed chiral density $n_5=n_R-n_L$ on timescales where chirality-changing processes are exhausted~\cite{andrianov_a,andrianov_b,andrianov_c,andrianov_d,andrianov_e,andrianov_f}. The Lagrangian is also invariant under the electromagnetic $U_Q(1)$ with $Q={\rm diag}(2/3,-1/3)$.

The motivation for retaining all four chemical potentials is astrophysical. Cold, dense, isospin-asymmetric quark matter is expected in neutron-star cores; most existing analyses keep only $\mu_B$ and $\mu_I$~\cite{son_a,son_b,son_c,he_a,he_b,he_c,he_d,Liu:2025dpw,WuZong:2017,ak_a,ak_b,ak_c,ak_d,ak_e,ekkz2_a,ekkz2_b,Andersen:2018nzq_a,Ayala:2024sqm,Lopes:2025rvn,XavierdeAzeredo:2026wlq,Ayala:2023mms,Basta:2025svw,Brandt:2024dle,Andersen:2023ivj,Bragutasymmetry,BragutaK,BragutaKS,Andersen:2018nzq_b,Ayala}. However, the strong magnetic fields characteristic of neutron stars induce nonzero $n_5$ and $n_{I5}$ via the chiral magnetic~\cite{fukus} and chiral separation~\cite{Metlitski} effects (see also~\cite{kkz19,Khun}). Including all four potentials therefore provides a more realistic framework for magnetised dense quark matter.

\subsection{Thermodynamic potential} \label{aaa1}

Before studying the thermodynamics, let us consider the partition function $\cal Z$ of the model (\ref{1}) as the following path integral:
\begin{eqnarray}
{\cal Z}_{\cal M}={\cal N}^{-1}\int {\cal D}\overline\psi {\cal D}\psi\exp\left\{i\int d^4x L_{NJL}\right\},
\label{a1}
\end{eqnarray}
where $\cal N$ is an infinite normalization constant and ${\cal D}\overline\psi{\cal D}\psi\equiv {\cal D}\overline\psi_u{\cal D}\overline\psi_d{\cal D}\psi_u{\cal D}\psi_d$. Note that if $\psi (x)\to U\psi (x)$ and $\overline\psi (x)\to \overline\psi (x) U^\dagger$, where $U$ is a 2$\times$2 flavor matrix, then the measures ${\cal D}\psi$ and ${\cal D}\overline\psi$ of the continual integral (\ref{a1}) are transformed according to the rule ((see, e.g., in Chapter 19 in Ref. \cite{Peskin} or in Chapter 22 in Ref. \cite{Weinberg}))
\begin{eqnarray}
{\cal D}\psi\to {\cal D}(U\psi)=(\det U)^{-1} {\cal D}\psi,~~{\cal D}\overline\psi\to {\cal D}(\overline\psi U^\dagger)=(\det U^\dagger)^{-1} {\cal D}\overline\psi.
\label{a4}
\end{eqnarray}
Hence, if $U\in U_{I_3}(1)$, i.e. $U=\cos\alpha
+\mathrm{i}\tau_3\sin\alpha$ (see in footnote \ref{f1,1}), then  $U U^\dagger=1$, and the measure ${\cal D}\overline\psi{\cal D}\psi$ of the path integral (\ref{a1}) remains intact.
But if $U\in U_{AI_3}(1)$, i.e. $U=\cos\alpha
+\mathrm{i}\gamma^5\tau_3\sin\alpha$ (see in footnote \ref{f1,1}), then $\psi_u\to e^{i\gamma^5\alpha}\psi_u$, $\psi_d\to e^{-i\gamma^5\alpha}\psi_d$ and $\overline\psi_u\to\overline\psi_u e^{i\gamma^5\alpha}$, $\overline\psi_d\to \overline\psi_de^{-i\gamma^5\alpha}$. In this case the path integral measure, ${\cal D}\overline\psi{\cal D}\psi\equiv {\cal D}\overline\psi_u{\cal D}\overline\psi_d{\cal D}\psi_u{\cal D}\psi_d$, transforms in the following way
\begin{eqnarray}
&&{\cal D}\overline\psi_u{\cal D}\overline\psi_d{\cal D}\psi_u{\cal D}\psi_d\to
{\cal D}(\overline\psi_u e^{i\gamma^5\alpha}){\cal D}(\overline\psi_d e^{-i\gamma^5\alpha}){\cal D}(e^{i\gamma^5\alpha}\psi_u ){\cal D}(e^{-i\gamma^5\alpha}\psi_d)
\nonumber\\&=& \big[\det(e^{2i\gamma^5\alpha})\det(e^{-2i\gamma^5\alpha})\big]^{-1}{\cal D}\overline\psi_u{\cal D}\overline\psi_d{\cal D}\psi_u{\cal D}\psi_d={\cal D}\overline\psi_u{\cal D}\overline\psi_d{\cal D}\psi_u{\cal D}\psi_d,\label{a5}
\end{eqnarray}
i.e. it is invariant under transformations from $U_{AI_3}(1)$ group.
Taking into account the trivial fact of the invariance of this measure with respect to $U_B(1)$ transformations, we can conclude that not only the Lagrangian $L_{NJL}$ (\ref{1}), but also the integration measure of the continual integral (\ref{a1}) does not change under transformations (\ref{2}). This means that the dynamics of dense quark matter, based on the Lagrangian (\ref{1}), is indeed invariant with respect to $U_B(1)\times U_{I_3}(1)\times U_{AI_3}(1)$ group.

Further, similar to the Hubbard-Stratonovich transformation used in Refs 
\cite{Gross,Eguchi,Coleman}, in our case also it is possible to introduce auxiliary scalar 
fields $\sigma(x)$ and $\pi_{1,2,3}(x)$ and then rewrite the partition function (\ref{a1}) in the 
following form 
\begin{eqnarray}
{\cal Z}_{\cal M}={\cal N'}^{-1}\int {\cal D}\overline\psi {\cal D}\psi {\cal D}\sigma {\cal D}\pi_1 {\cal D}\pi_2 {\cal D}\pi_3\exp\left\{i\int d^4x \widetilde L_{NJL}\right\},
\label{a2}
\end{eqnarray}
where 
\begin{eqnarray}
\widetilde L_{NJL}
&=&\overline \psi\Big [\gamma^\rho\mathrm{i}\partial_\rho
-\sigma
-\mathrm{i}\gamma^5\vec\pi\cdot\vec\tau\Big ]\psi+\overline{\psi}\gamma^0{\cal M}\psi
 -\frac{N_c}{4G}\Big [\sigma^2+\vec\pi^2\Big ]\nonumber\\
&=&L_{NJL}-\frac{N_c}{4G}\Big[\sigma+2\frac{G}{N_c}(\overline\psi\psi)\Big]^2-
\frac{N_c}{4G}\sum_{a=1}^3\Big[\pi_a+2\frac{G}{N_c}(i\overline\psi\gamma^5\tau_a\psi)\Big]^2.
\label{3}
\end{eqnarray}
It is obvious that after integration in Eq. (\ref{a2}) over the fields $\sigma$ and 
$\vec\pi$, we obtain (\ref{a1}). It is easy to see from Eq. (\ref{3}) that  $\widetilde L_{NJL}$ differs from the original Lagrangian $L_{NJL}$ by two terms, which however have no effects on the dynamics of the model (\ref{1}).
It is clear from the fact that the Euler-Lagrange equations for auxiliary scalar fields are  
\begin{eqnarray}
\sigma(x)=-2\frac{ G}{N_c}(\overline\psi\psi);~~~\pi_a (x)=-2\frac{ G}{N_c}(\overline\psi\mathrm{i}\gamma^5\tau_a\psi).
\label{4}
\end{eqnarray}
These relations contain no time derivatives and therefore do not represent genuine equations of motion but rather algebraic constraint equations. Consequently, when constructing the Hamiltonian of the system based on the Lagrangian $\widetilde L_{NJL}$, one must take these constraints~(\ref{4}) into account. As is evident from Eq.~(\ref{3}), once the constraints are imposed, $\widetilde L_{NJL}$ reduces exactly to the original Lagrangian~$L_{NJL}$. This means that the dynamics defined by the two Lagrangians are identical, although their corresponding Feynman rules differ.  

The advantage of the bosonized Lagrangian $\widetilde L_{NJL}$ lies in the presence of auxiliary scalar fields, which describe collective quark states in different interaction channels. Owing to this feature, $\widetilde L_{NJL}$ rather than the original $L_{NJL}$ is more convenient for analyzing the properties of the ground state, i.e. the phase structure of dense quark matter. To this end, one proceeds from the partition function~(\ref{a2}) to the generating functional ${\cal Z}_{\cal M}(J)$ of the Green's functions of the scalar fields,
\begin{eqnarray}
{\cal Z}_{\cal M}(J)=\exp^{iN_c{\cal W}_{\cal M}(J)}={\cal N'}^{-1}\int {\cal D}\overline\psi 
{\cal D}\psi {\cal D}\sigma \prod_{i=1}^3{\cal D}\pi_i\exp\left\{i\int d^4x\big [ \widetilde 
L_{NJL}+N_c(\sigma J_0+\vec\pi\cdot\vec J)\big]\right\},
\label{a3}
\end{eqnarray}
where $\vec\pi\cdot\vec J\equiv \pi_1 J_1+\pi_2J_2+\pi_3J_3$ and $J_0(x)$ and $J_{1,2,3}(x)$ are scalar sources. 

Let us assume that the spinor fields are transformed according to one of the Abelian groups 
(\ref{2}). Let us find how the scalar fields $\sigma (x)$, $\pi_a(x)$ should be transformed 
so that in this case the auxiliary Lagrangian (\ref{3}) remains invariant. First, it is 
evident that if spinor doublet $\psi$ is transformed by $U_B(1)$ group, then for this it is 
sufficient that the auxiliary scalar fields remain intact. Second, if spinor doublet is 
transformed by $U_{I_3}(1)$ or $U_{AI_3}(1)$, then, taking into account the footnote 
\ref{f1,1} and the explicit form of $\widetilde L_{NJL}$  (\ref{3}), it is possible to show
that this Lagrangian will not be changed, if the auxiliary scalar fields are transformed, respectively, according to the following rules
\begin{eqnarray}
U_{I_3}(1):~&&\sigma\to\sigma;~~\pi_3\to\pi_3;~~\pi_1\to\cos
(\alpha)\pi_1+\sin (\alpha)\pi_2;~~\pi_2\to\cos (\alpha)\pi_2-\sin
(\alpha)\pi_1,\label{50}
\\
U_{AI_3}(1):~&&\pi_1\to\pi_1;~~\pi_2\to\pi_2;~~\sigma\to\cos
(\alpha)\sigma+\sin (\alpha)\pi_3;~~\pi_3\to\cos
(\alpha)\pi_3-\sin (\alpha)\sigma.
\label{5}
\end{eqnarray}
Note that the path integral (\ref{a3}) measure ${\cal D}\sigma \prod_{i=1}^3{\cal D}\pi_i$ is invariant 
under these transformations of scalar fields.
Whether symmetry (\ref{50})-(\ref{5}) breakings occurs or not can be deduced from the 
thermodynamic potential (TDP). To get the TDP, we first obtain the effective action 
$\Gamma (\sigma_{cl},\pi_{a_{cl}})$ using the Legendre transform for the functional ${\cal W}_{\cal M}(J)$ (\ref{a3}):  
\begin{eqnarray}
\Gamma (\sigma_{cl},\pi_{a_{cl}})={\cal W}_{\cal M}(J)-\int d^4x\left\{\sigma_{cl}(x)J_0(x)+\sum_{a=1}^3
\pi_{a_{cl}}(x)J_a(x)\right\},\label{a6}
\end{eqnarray}
where $\sigma_{cl} (x)=\delta{\cal W}_{\cal M}(J)/\delta J_0(x)$ and 
$\pi_{a_{cl}} (x)=\delta{\cal W}_{\cal M}(J)/\delta J_a(x)$ are the ground state expectation 
values of the auxiliary scalar fields in the presence of external sources, i.e. $\sigma_{cl} (x)=
\langle\sigma(x)\rangle_J$ and $\pi_{a_{cl}} (x)=\langle\pi_a(x)\rangle_J$ (see Eq. (\ref{aa1}) 
in Appendix A). The classical fields $\sigma_{cl} (x)$ and $\pi_{a_{cl}} (x)$ obey the following equations
\begin{eqnarray}
\frac{\delta\Gamma_{{\cal M}} (\sigma_{cl},\pi_{a_{cl}})}{\delta\sigma_{cl}(x)}=-J_0(x),~~~~
\frac{\delta\Gamma_{{\cal M}} (\sigma_{cl},\pi_{a_{cl}})}{\delta\pi_{k_{cl}}(x)}=-J_k(x).\label{a7}
\end{eqnarray}
From this it is clear that expectation values $\langle\sigma (x)\rangle$ and 
$\langle\pi_{1,2,3} (x)\rangle$ of auxiliary scalar fields over the truly ground state 
of the system, must satisfy the equations (\ref{a7}) with zero external sources.
In addition, we assume that our physical system is invariant under spacetime translations. It
means that $\langle\sigma (x)\rangle$ and $\langle\pi_{1,2,3} (x)\rangle$ should not depend on coordinates. So these quantities are the components of the global minimum point of the TDP $\Omega_{{\cal M}}(\sigma_{cl},\pi_{a_{cl}})$ which is defined by the relation
\begin{eqnarray}
\Gamma_{{\cal M}} (\sigma_{cl},\pi_{a_{cl}})\Big |_{\sigma_{cl},\pi_{a_{cl}}=const}
=-\Omega_{{\cal M}}(\sigma_{cl},\pi_{a_{cl}})\int d^4x.\label{7}
\end{eqnarray} 
Then phase structure of the NJL model (\ref{1}) is determined by the behaviour of $\langle\sigma (x)\rangle$ and $\langle\pi_{1,2,3} (x)\rangle$ vs chemical potentials. And the problem is significantly simplified since, as it is shown in Appendix A, the TDP $\Omega_{{\cal M}}(\sigma_{cl},\pi_{a_{cl}})$ is really a function of 
only two field combinations, $\Sigma\equiv\sqrt{\sigma^2_{cl}+\pi_{3_{cl}}^2}$ and 
$\Pi\equiv\sqrt{\pi_{1_{cl}}^2+\pi_{2_{cl}}^2}$, i.e.
\begin{eqnarray}
\Omega_{{\cal M}}(\sigma_{cl},\pi_{a_{cl}})\equiv \Omega \big(\Sigma,\Pi\big).
\label{8}
\end{eqnarray}
 
\subsection{Dual symmetry in the mean-field approximation} \label{aaa2}

For simplicity, both the phase structure of the NJL model (\ref{1}) and its TDP (\ref{7}) is
usually studied in the so-called mean-field approximation or, equivalently, in the leading 
order of the large-$N_c$ expansion technique. In this case, using a general formula   
\begin{eqnarray}
{\cal N'}^{-1}\int {\cal D}\overline\psi 
{\cal D}\psi \exp\left\{i\int d^4x\overline\psi D\psi\right\}=[\det D]^{N_c},
\label{a8}
\end{eqnarray}
one can integrate in Eq. (\ref{a3}) over spinor fields. Then
\begin{eqnarray}
{\cal Z}_{\cal M}(J)=\exp^{iN_c{\cal W}_{\cal M}(J)}={\cal N'}^{-1}\int {\cal D}\sigma 
\prod_{k=1}^3{\cal D}\pi_k
\exp\left\{iN_c \left(S_{eff}(\sigma,\pi_a)+\int d^4x\big [ \sigma J_0+\vec\pi\cdot\vec J
\big]\right)\right\},
\label{a9}
\end{eqnarray} where 
\begin{eqnarray}
{\cal S}_{\rm {eff}}(\sigma(x),\pi_a(x))
=-\int
d^4x\left[\frac{\sigma^2(x)+\pi^2_a(x)}{4G}\right]-\mathrm{i}{\rm
Tr}_{sfx}\ln D,\nonumber\\
D=\gamma^\nu\mathrm{i}\partial_\nu +\frac{\mu_B}{3}\gamma^0
+ \frac{\mu_I}{2}\tau_3\gamma^0+\frac{\mu_{I5}}{2}\tau_{3}\gamma^0\gamma^5+\mu_{5}\gamma^0\gamma^5-\sigma (x) -\mathrm{i}\gamma^5\pi_a(x)\tau_a.
\label{a10}
\end{eqnarray}
Here the Tr-operation stands for the trace in spinor- ($s$), flavor-
($f$) as well as four-dimensional coordinate- ($x$) spaces, respectively. As it is shown in Appendix \ref{ApB}, 
starting from Eq. (\ref{a9}) one can easily find effective action $\Gamma_{mf}(\sigma_{cl}(x),\pi_{a_{cl}}(x))$ of the model in the leading large-$N_c$ order (or in the mean-field approximation),
\begin{eqnarray}
\Gamma_{mf}(\sigma_{cl}(x),\pi_{a_{cl}}(x))=
{\cal S}_{\rm {eff}}(\sigma_{cl}(x),\pi_{a_{cl}}(x)).\label{a11}
\end{eqnarray}
It follows from Eq. (\ref{7}) that the mean-field approximation for the TDP of the model can be found by the relation 
\begin{eqnarray}
\Gamma_{mf}(\sigma_{cl}(x),\pi_{a_{cl}}(x))\Big |_{\sigma_{cl}(x),\pi_{a_{cl}}(x)=const}
=-\Omega_{mf}(\sigma_{cl},\pi_{a_{cl}})\int d^4x.\label{a12}
\end{eqnarray}
In the recent paper \cite{kkz18-2} the mean-field TDP (\ref{a12}) has been investigated. 
In particular, it was shown there that $\Omega_{mf}(\sigma_{cl},\pi_{a_{cl}})$ is indeed a 
function of only two variables $\Sigma$ and $\Pi$ (these quantities are defined in the text
before Eq. (\ref{8})), i.e.  
$\Omega_{mf}(\sigma_{cl},\pi_{a_{cl}})\equiv\Omega_{mf}(\Sigma,\Pi)$, and the mean-field TDP is invariant under the transformation
\begin{eqnarray}
\widetilde{\cal D}:~~\Sigma \longleftrightarrow\Pi,~~~\mu_{I} 
\longleftrightarrow \mu_{I5}.\label{9}
\end{eqnarray}
Due to this property,
it was shown in Ref. \cite{kkz18-2} that
there is a duality between CSB and charged PC phenomena (so we call the $\widetilde{\cal D}$ (\ref{9}) as a dual transformation).
It means that at fixed $\mu_B$ and $\mu_5$ the $(\mu_I,\mu_{I5})$-phase portrait (obtained in the framework of the 
mean-field approximation) of the NJL model (\ref{1}) obeys a symmetry with
respect to simultaneous transformations, CSB$\leftrightarrow$charged PC and $\mu_{I}\leftrightarrow\mu_{I5}$.

In more detail, suppose that for certain fixed values of the chemical potentials $\mu_B$, $\mu_{5}$, $\mu_I=A$, and $\mu_{I5}=B$, the global minimum of the thermodynamic potential $\Omega_{mf}(\Sigma,\Pi)$ lies at the point $(\Sigma=\Sigma_0\neq0,\Pi=0)$. In this case, for the chosen set of chemical potentials, the chiral $U_{AI_3}(1)$ symmetry (\ref{5}) is spontaneously broken, and the model is realized in the chiral symmetry breaking (CSB) phase.  

Then, due to the invariance of the thermodynamic potential with respect to the dual transformation $\widetilde{\cal D}$ (\ref{9}), it immediately follows that at the interchanged values of the chemical potentials, $\mu_I=B$ and $\mu_{I5}=A$, while keeping $\mu_B$ and $\mu_{5}$ unchanged, the global minimum of $\Omega_{mf}(\Sigma,\Pi)$ is located at the point $(\Sigma=0,\Pi=\Sigma_0)$. As can be seen from Eq.~(\ref{50}), in this region of the phase diagram the isospin $U_{I_3}(1)$ symmetry is spontaneously broken, and the system resides in the charged pion condensation (PC) phase.  

Consequently, knowledge of the phase realized in the model (\ref{1}) for a particular set of external parameters $\mu_B$, $\mu_I$, $\mu_{I5}$, and $\mu_5$ is sufficient to determine the corresponding \emph{dually conjugated} phase realized at the rearranged values $\mu_I\leftrightarrow\mu_{I5}$, with $\mu_B$ and $\mu_5$ kept fixed.  

Moreover, various physical quantities---such as condensates, densities, and other observables---that characterize the two dually related phases are connected by the same dual transformation $\widetilde{\cal D}$. For example, the chiral condensate in the CSB phase at given $\mu_B$, $\mu_I$, $\mu_{I5}$, and $\mu_5$ equals the charged pion condensate in the corresponding dually conjugated PC phase, where one performs the interchange $\mu_I\leftrightarrow\mu_{I5}$. Similarly, knowing the quark number density $n_q(\mu_I,\mu_{I5})$ in the initial CSB phase allows one to obtain the density in the conjugated PC phase simply by swapping the arguments, $n_q(\mu_I,\mu_{I5})\to n_q(\mu_{I5},\mu_I)$, and the same applies to all other thermodynamic quantities.  

Thus, the symmetry of the thermodynamic potential of the massless NJL model (\ref{1}) with respect to the dual transformation (\ref{9}) greatly simplifies the analysis of dense ($\mu_B\neq0$) quark matter with simultaneous isospin ($\mu_I\neq0$) and chiral isospin ($\mu_{I5}\neq0$) asymmetries. However, it should be emphasized once again that the duality between the CSB and charged PC phenomena, established in Ref.~\cite{kkz18-2}, was demonstrated only at the level of the thermodynamic potential calculated within the mean-field (leading-order $1/N_c$) approximation. Whether this duality persists beyond mean-field, i.e. in higher orders of the $1/N_c$ expansion, or holds for the full NJL model itself, remains an open and nontrivial question.

\section{Dual symmetry of the simplest NJL Lagrangian}\label{aaa3}

In the present paper we demonstrate that: (i) there exists a discrete Pauli--Gursey-type transformation of the quark fields\footnote{The most general form of the Pauli--Gursey transformation is given below Eq.~(\ref{17}).} that converts the Lagrangian~(\ref{1}) term responsible for the chiral symmetry breaking (CSB) interaction channel into the corresponding term describing the interaction channel where charged pion condensation occurs, and vice versa.  

Furthermore, (ii) under the same transformation, the Lagrangian~(\ref{1}) terms corresponding to the isospin and chiral isospin charge densities are interchanged, while all other terms of the NJL Lagrangian remain unaffected.  

Consequently, (iii) if the interchange $\mu_I\leftrightarrow\mu_{I5}$ is performed simultaneously with this Pauli--Gursey-type transformation, the entire Lagrangian~(\ref{1}) remains invariant. As a result, the dual symmetry~(\ref{9}) is a property of the full thermodynamic potential $\Omega(\Sigma,\Pi)$ of the model, and not merely of its mean-field approximation $\Omega_{mf}(\Sigma,\Pi)$.
To study the problem, let us define the following transformation of the flavor doublet $\psi(x)$ (as it is clear from the following consideration, it is indeed a Pauli-Gursey transformation),
\begin{eqnarray}
PG:~~\psi_R\to\psi_R,~~~\psi_L\to i\tau_1\psi_L,~~ {\rm or}~~\overline{\psi_R}\to \overline{\psi_R},~~
\overline{\psi_L}\to \overline{\psi_L} (-i)\tau_1,\label{10}
\end{eqnarray}
where $\psi=\psi_L+\psi_R$ and
\begin{eqnarray*}
\psi_R\equiv\frac{1+\gamma^5}2\psi\equiv\Pi_+\psi,~~~\psi_L\equiv\frac{1-\gamma^5}2\psi\equiv\Pi_-\psi,~~{\rm i.e.}
~~\overline{\psi_R}=\overline\psi\Pi_-,~~\overline{\psi_L}=\overline\psi\Pi_+.\label{11}
\end{eqnarray*}
Alternatively, it can be presented in the following form
\begin{eqnarray}
PG:~~\psi\to \psi_R+i\tau_1\psi_L,~~~{\rm i.e.}~~~ PG\psi\equiv PG\left( {\begin{array}{c}
	 \psi_{u} \\
	 \psi_d
	\end{array} } \right)=\left( {\begin{array}{c}
	 \psi_{uR}+i\psi_{dL} \\
	 \psi_{dR}+i\psi_{uL}
	\end{array} } \right).\label{12}
\end{eqnarray}
Using Eqs. (\ref{10})-(\ref{12}), it is easy to find out how the simplest quark-antiquark 
structures of the Lagrangian (\ref{1}) are transformed under the action of this 
Pauli--Gursey-type transformation,
\begin{eqnarray}
\overline{\psi}\psi=\overline{\psi_{R}}\psi_{L}+\overline{\psi_{L}}\psi_{R}&\stackrel{PG}{\longrightarrow}&
\overline{\psi_{R}}i\tau_{1}\psi_{L}-
\overline{\psi_{L}}i\tau_{1}\psi_{R}=-i \overline{\psi}\gamma^{5}\tau_{1}\psi\label{13}\\
i\overline{\psi}\gamma^{5}\tau_{1}\psi=i(-\overline{\psi_{R}}\tau_{1}\psi_{L}+
\overline{\psi_{L}}\tau_{1}\psi_{R})
&\stackrel{PG}{\longrightarrow}&
-i\overline{\psi_{R}}\tau_{1}i\tau_{1}\psi_{L}-i\overline{\psi_{L}}i\tau_{1}\tau_{1}\psi_{R}=\overline{\psi}\psi.
\label{14}
\end{eqnarray}
In a similar way it is possible to show that
\begin{eqnarray}
i \overline{\psi}\gamma^{5}\tau_{2}\psi
&\stackrel{PG}{\longrightarrow}&
i\overline{\psi}\gamma^{5}\tau_{3}\psi,\label{15}\\
i \overline{\psi}\gamma^{5}\tau_{3}\psi 
&\stackrel{PG}{\longrightarrow}&
-i\overline{\psi}\gamma^{5}\tau_{2}\psi.\label{16}
\end{eqnarray}
It means that four-fermion structures of the Lagrangian (\ref{1}) that are responsible for the CSB and charged PC
are transformed to each other by the $PG$ transformation (\ref{10})-(\ref{12}), i.e.
\begin{equation}
(\overline{\psi}\psi)^2+
(i\overline{\psi}\gamma^5\tau_3\psi)^2\stackrel{PG}{\longleftrightarrow} (i\overline{\psi}\gamma^5\tau_1\psi)^2
+(i\overline{\psi}\gamma^5\tau_2\psi)^2. \label{17}
\end{equation}
(Note that original Pauli--Gursey transformation of Ref. \cite{pauli_a,pauli_b} connects the four-fermion structures of 
low-dimensional Lagrangians responsible for the CSB and superconducting channels (see, e.g., in Refs. 
\cite{ebert,cao,ekkz,ojima}). So, in order to highlight this particular feature of the 
Pauli--Gursey transformation, in the present paper {\bf we call any transformation of spinor fields 
that connects the quark structures of various interaction channels to each other as a 
Pauli--Gursey-type one.})
Since the free term $i\overline{\psi} \gamma^{\mu}\partial_{\mu} \psi$ of the Lagrangian (\ref{1}) remains intact under the action of the $PG$ transformation,
it is clear that at zero chemical potentials the initial NJL Lagrangian (\ref{1}) is invariant under the Pauli--Gursey-type
transformation (\ref{10})-(\ref{12}). However, at nonzero chemical potentials it is no more the symmetry
transformation of the NJL model. Indeed, under the action of $PG$ transformation (\ref{10})-(\ref{12}) the quantities
$\overline{\psi}\gamma^{0}\psi$ and $\overline{\psi}\gamma^{0}\gamma^{5}\psi$ are not changed, but
\begin{eqnarray}
\overline{\psi}\gamma^{0}\gamma^{5}\tau_{3}\psi&=&(\overline{\psi_{R}}\gamma^{0}\tau_{3}\psi_{R}-
\overline{\psi_{L}}\gamma^{0}\tau_{3}\psi_{L})\stackrel{PG}{\longleftrightarrow}(\overline{\psi_{R}}\gamma^{0}\tau_{3}\psi_{R}+
\overline{\psi_{L}}i\tau_{1}\gamma^{0}\tau_{3}i\tau_{1}\psi_{L})\nonumber\\&=&(\overline{\psi_{R}}\gamma^{0}\tau_{3}\psi_{R}+
\overline{\psi_{L}}\gamma^{0}\tau_{3}\psi_{L})=\overline{\psi}\gamma^{0}\tau_{3}\psi.\label{18}
\end{eqnarray}

It follows that the densities of the isospin and chiral isospin charges, i.e. the quantities 
$\overline{\psi}\gamma^{0}\tau_{3}\psi$ and $\overline{\psi}\gamma^{0}\gamma^{5}\tau_{3}\psi$ 
from Eq.~(\ref{18}), are transformed into one another. 
As a result, when both $\mu_I\ne0$ and $\mu_{I5}\ne0$, 
the NJL Lagrangian~(\ref{1}) is no longer invariant under the Pauli--Gursey (PG) transformation. 
Hence, at nonvanishing $\mu_I$ and $\mu_{I5}$, the discrete PG transformation ceases to be a true symmetry of the model.  

Here $\mu_I$ and $\mu_{I5}$ act as external classical backgrounds coupled to the isospin and
chiral isospin currents $\overline\psi\gamma^0\tau_3\psi$ and $\overline\psi\gamma^0\gamma^5\tau_3\psi$.
Since the $PG$ rotation swaps these two currents [Eq.~(\ref{18})], swapping their sources
$\mu_I\leftrightarrow\mu_{I5}$ is exactly what restores the symmetry and maps the two source terms onto each other. This can be also seen if we rewrite source terms into
$$
\frac{\mu_I}{2}\,\overline\psi\gamma^0\tau_3\psi+\frac{\mu_{I5}}{2}\,\overline\psi\gamma^0\gamma^5\tau_3\psi
=\frac{\mu_I+\mu_{I5}}{2}\,\overline{\psi_R}\gamma^0\tau_3\psi_R
+\frac{\mu_I-\mu_{I5}}{2}\,\overline{\psi_L}\gamma^0\tau_3\psi_L.
$$
The $PG$ rotation~(\ref{10}) touches only the left-handed doublet, $\psi_L\to i\tau_1\psi_L$, and
flips the sign of its bilinear, $\overline{\psi_L}\gamma^0\tau_3\psi_L\to-\overline{\psi_L}\gamma^0\tau_3\psi_L$
(since $\tau_1\tau_3\tau_1=-\tau_3$), while leaving the right-handed one intact. It thus induces a minus sign to
$(\mu_I-\mu_{I5})$ term and keeps $(\mu_I+\mu_{I5})$ intact, and the interchange $\mu_I\leftrightarrow\mu_{I5}$
is precisely what compensates the sign. 

However, it is evident that if the permutation of the chemical potentials 
$\mu_I\longleftrightarrow\mu_{I5}$ is performed simultaneously with the field transformations~(\ref{10})--(\ref{12}), 
the NJL Lagrangian remains invariant. 
\textbf{Transformations under which not only the field variables of the Lagrangian but also some of the external 
parameters of the system (such as chemical potentials, coupling constants, etc.) are simultaneously transformed will hereafter be referred to as \emph{dual transformations}.}  
Thus, the NJL Lagrangian~(\ref{1}) is invariant (or symmetric) under the discrete dual transformation ${\cal D}$ defined as
\begin{eqnarray}
{\cal D}:~~\psi\to \psi_R+i\tau_1\psi_L,~~~\mu_I\longleftrightarrow\mu_{I5}.
\label{20}
\end{eqnarray}
In this case, the quark bilinears corresponding to the CSB and charged PC interaction channels transform into one another 
(see, e.g., Eq.~(\ref{17})). Therefore, we say that these channels are ${\cal D}$-dual to each other, 
or equivalently, that the CSB and charged PC phenomena are \emph{dually ${\cal D}$ conjugated} within the framework 
of the NJL model~(\ref{1}).

Recall that the full thermodynamic potential (TDP)~(\ref{7})--(\ref{8}) of the NJL model is derived from the auxiliary 
bosonized Lagrangian $\widetilde L_{NJL}$~(\ref{3}). 
It is therefore natural to ask how the scalar fields $\sigma(x)$ and $\vec\pi(x)$ should transform under the dual 
transformation ${\cal D}$~(\ref{20}) of the quark fields so that the auxiliary Lagrangian $\widetilde L_{NJL}$ 
remains invariant as a whole. 
Taking into account that the quark bilinears in $\widetilde L_{NJL}$ transform according to Eqs.~(\ref{13})--(\ref{16}), 
it is straightforward to verify that the auxiliary scalar fields must transform as
\begin{eqnarray}
PG,~{\cal D}:~~\sigma(x)\to -\pi_1(x),~~\pi_1(x)\to \sigma(x),~~
\pi_2(x)\to \pi_3(x),~~\pi_3(x)\to -\pi_2(x),
\label{21}
\end{eqnarray}
in order to leave $\widetilde L_{NJL}$ invariant. 
The same transformation rules~(\ref{21}) can also be obtained directly from the Euler--Lagrange equations~(\ref{4}) for 
the auxiliary fields if one performs the dual transformation~(\ref{20}) of the quark fields therein.

Importantly, Appendix~C provides a direct proof of the invariance of the functional measure 
${\cal D}\overline\psi{\cal D}\psi$ in the partition function~(\ref{a1}) under the PG (and hence dual) transformation~(\ref{20}). 
As a consequence, the full effective action $\Gamma_{\cal M}(\sigma_{cl},\vec\pi_{cl})$ and the full thermodynamic potential 
$\Omega_{\cal M}(\sigma_{cl},\vec\pi_{cl})$~(\ref{8}) are invariant under the dual transformations 
(\ref{C10})--(\ref{C11}) of the classical fields $\sigma_{cl}(x)$ and $\pi_{a\,cl}(x)$, accompanied by the interchange 
of chemical potentials $\mu_I\leftrightarrow\mu_{I5}$ (see, e.g., Eq.~(\ref{C13}) in Appendix~\ref{ApC}). 
This property---the ${\cal D}$-symmetry of the full TDP---is precisely the invariance of the mean-field TDP 
under the discrete dual transformation $\widetilde{\cal D}$~(\ref{9}), corresponding to the substitutions 
$\Sigma\leftrightarrow\Pi$ and $\mu_I\leftrightarrow\mu_{I5}$. 
Physically, this symmetry manifests itself as the duality between the CSB and charged PC phenomena.

Therefore, we conclude that the dual symmetry between the CSB and charged PC phases of dense quark matter with 
isospin and chiral isospin asymmetries (as described below Eq.~(\ref{9}))---originally observed in Ref.~\cite{kkz18-2} 
within the mean-field approximation---is in fact an intrinsic property of the massless NJL model~(\ref{1}) itself. 
Hence, the duality between CSB and charged PC phases is realized throughout the full 
$(\mu_B,\mu_I,\mu_{I5},\mu_5)$ phase diagram of the model, and not only within the mean-field (leading $1/N_c$) framework. 
This follows directly from the dual symmetry ${\cal D}$~(\ref{20}) of both the microscopic NJL Lagrangian~(\ref{1}) and the 
functional integration measure ${\cal D}\overline\psi{\cal D}\psi$ in the path integral.

\section{Dual symmetry of NJL Lagrangian with color superconductivity }\label{aaa3a}

Let us consider more general type of NJL model. Recall that in the recent paper \cite{u2} the phase structure of the four-fermion model, whose Lagrangian $L_{CSC}$,
\begin{equation}
L_{CSC}=L_{NJL}+H\sum_{k=1,2,3}
[\overline{\psi^c}\gamma^5\tau_2\epsilon_k\psi]
[\overline{\psi} \gamma^5\tau_2\epsilon_k \psi^c],\label{100}
\end{equation}
differs from Eq. (\ref{1}) by an additional term with diquark interaction, was discussed in the mean-field approximation. 
In Eq. (\ref{100}),  
$\psi^c=C\overline\psi^T$, 
$\overline{\psi^c}=\psi^T C$ are charge-conjugated spinors, where $C=i\gamma^2\gamma^0$ is
the charge conjugation matrix (the symbol $T$ denotes the transposition operation), $\epsilon_k$ ($k=1,2,3$) are 3$\times$3 matrices acting in the three-dimensional fundamental representation of the color $SU_c(3)$ group, where they have matrix elements $(\epsilon_k)_{ij}=\epsilon_{ijk}$ (the last quantity is the totally antisymmetric tensor over indices $i,j,k=1,2,3$). Other notations are the same 
as for the Lagrangian $L_{NJL}$ (see the text below Eq. (\ref{1})). As in the case with the simpler NJL model from 
the previous section II, when considering the phase structure of the model (\ref{100}) it is
also convenient to deal with its auxiliary Lagrangian $\widetilde L_{CSC}$, 
\begin{eqnarray}
\widetilde L_{CSC}
&=&\widetilde L_{NJL}-\frac1{4H}\Delta^{*}_{k}\Delta_{k}-
 \frac{\Delta^{*}_{k}}{2}[\overline{\psi^c}\gamma^5\tau_2\epsilon_k \psi]
-\frac{\Delta_{k}}{2}[\overline\psi \gamma^5\tau_2\epsilon_{k}\psi^c],
\label{CSC}
\end{eqnarray}
where $\widetilde L_{NJL}$ is an auxiliary semibosonized Lagrangian (\ref{3}). In addition to
spinor fields $\psi(x)$ and auxiliary bosonic fields $\sigma(x),\vec\pi(x)$ (see in Eq. (\ref{4})), 
$\widetilde L_{CSC}$ contains auxiliary scalar 
diquark fields $\Delta_k(x)$ and $\Delta^*_{k'}(x)$, where $k,k'=1,2,3$, which transform as color 
$SU_c(3)$ antitriplet and triplet, respectively. Moreover, the Euler-Lagrange equations of motion for 
these fields look like  
\begin{eqnarray}
\Delta_k(x)=-2H[\overline{\psi^c}\gamma^5\tau_2\epsilon_{k}\psi],~~\Delta^*_k(x)=
-2H[\overline{\psi}\gamma^5\tau_2\epsilon_{k}\psi^c].\label{99}
\end{eqnarray}
Their ground state
expectation values are the order parameters of the color superconducting phase. \footnote{The details of the construction of the auxiliary 
semi-bosonic Lagrangian $\widetilde L_{CSC}$ 
are described  both in the paper \cite{u2} and in earlier reviews devoted to color superconductivity \cite{buballa_a,buballa_b,buballa_c,buballa_d,buballa_e,buballa_f}.} 

On the basis of the microscopic auxiliary Lagrangian (\ref{CSC}), it is much easier to 
construct formally effective action $\Gamma_{CSC} (\sigma_{cl},\pi_{a_{cl}},\Delta_{k_{cl}},\Delta^*_{k'_{cl}})$.  
First, one can write down the generating functional ${\cal Z}_{CSC}(J,\overline{\eta},\eta)$ of the model (\ref{100}),
\begin{eqnarray}
{\cal Z}_{CSC}(J, \overline{\eta}, \eta )={\cal N'}^{-1}\int {\cal D}\overline\psi 
{\cal D}\psi {\cal D}\sigma \prod_{i=1}^3{\cal D}\pi_i{\cal D}\Delta_i{\cal D}\Delta^*_i\exp\left\{i\int d^4x\big [ \widetilde 
L_{CSC}+N_c(\sigma J_0+\vec\pi\cdot\vec J+\overline{\eta}_k\Delta_k+\eta_k\Delta^*_k  )\big]\right\},\label{101}
\end{eqnarray}
One can construct effective action using generating functional ${\cal W}_{CSC}(J, \overline{\eta} , \eta ):{\cal Z}_{CSC}(J, \overline{\eta} , \eta )=\exp^{iN_c{\cal W}_{CSC}(J, \overline{\eta}, \eta )}$ and
$$\Gamma_{CSC}(\sigma_{cl},\pi_{a_{cl}}, \Delta_{k_{cl}},\Delta^*_{k'_{cl}})={\cal W}_{CSC}(J, \overline{\eta} , \eta )-\int d^4x\left\{\sigma_{cl}(x)J_0(x)+
\pi_{a_{cl}}(x)J_a(x)+\overline{\eta}_a(x) \Delta_{a_{cl}}(x)+\eta_a(x) \Delta^*_{a_{cl}}(x)  \right\},$$
where $N_c=3$ and summation over the repeated indices $a=1,2,3$ and $k=1,2,3$ (in Eq. (\ref{101}))  
is implied. The arguments of the effective action, i.e. the classical scalar fields 
$\sigma_{cl},\pi_{a_{cl}}, \Delta_{k_{cl}},\Delta^*_{k'_{cl}}$, are defined by the following equations
\begin{eqnarray}
\sigma_{cl} (x)=\frac{\delta{\cal W}_{CSC}}{\delta J_0(x)},\;\;\;\;\;\pi_{a_{cl}} (x)=
\frac{\delta{\cal W}_{CSC}}{\delta J_a(x)},\;\;\;\;\;\Delta_{k_{cl}}(x)=\frac{\delta{\cal W}_{CSC}}{\delta\, 
\overline{\eta}_k(x)},\;\;\;\;\; \Delta^*_{k_{cl}}(x)=\frac{\delta{\cal W}_{CSC}}{\delta \eta_k(x)},
\end{eqnarray}
from which it is clear (see also Appendix \ref{ApA}) that classical scalar fields are the ground 
state expectation values of the auxiliary scalar fields in the presence of external sources, i.e. 
$\sigma_{cl}(x)=\langle\sigma(x)\rangle_J$, $\Delta_{k_{cl}}(x)=\langle\Delta_{k}(x)\rangle_\eta$,...

One can easily show that the fields $\sigma_{cl} (x)$, $\pi_{a_{cl}} (x)$, $\Delta_{k_{cl}} (x)$  and $\Delta^*_{k'_{cl}}(x)$ should obey the following equations
\begin{eqnarray}
\frac{\delta\Gamma_{CSC}}{\delta\sigma_{cl}(x)}=-J_0(x),~~~~~~
\frac{\delta\Gamma_{CSC} }{\delta\pi_{a_{cl}}(x)}=-J_a(x),\;\;\;\;\;\frac{\delta\Gamma_{CSC} }{\delta\Delta_{k_{cl}}(x)}=- \overline{\eta}_k(x),\;\;\;\;\;\frac{\delta\Gamma_{CSC} }{\delta\Delta^*_{k_{cl}}(x)}=- \eta_k(x).
\label{De8}
\end{eqnarray}
Similarly to the discussion in the above section II, expectation values $\langle\sigma (x)\rangle$,
$\langle\pi_{1,2,3} (x)\rangle$, $\langle\Delta_{1,2,3} (x)\rangle$  and $\langle\Delta^*_{1,2,3}(x)\rangle$ of auxiliary scalar fields over the truly ground state of the system must fulfill the equations (\ref{De8}) with zero external sources $J_0=J_a=\overline{\eta}_k=\eta_l=0$.
Also, our physical system is assumed to be invariant under spacetime translations which entails that $\langle\sigma (x)\rangle$, $\langle\pi_{1,2,3} (x)\rangle$, $\langle\Delta_k (x)\rangle$  and $\langle\Delta^*_l(x)\rangle$ should not depend on coordinates. So defining the TDP $\Omega_{CSC}(\sigma_{cl},\pi_{a_{cl}},\Delta_{k_{cl}},\Delta^*_{k'_{cl}})$ by the relation
\begin{eqnarray}
\Gamma_{CSC} (\sigma_{cl},\pi_{a_{cl}},\Delta_{k_{cl}},\Delta^*_{k'_{cl}})\Big |_{...,\Delta_{k_{cl}},\Delta^*_{k'_{cl}}=const}=-\Omega_{CSC}(\sigma_{cl},\pi_{a_{cl}},\Delta_{k_{cl}},\Delta^*_{k'_{cl}})\int d^4 x,
\label{TDP}
\end{eqnarray}
which is indeed a function of constant 
scalar fields $\sigma_{cl},\pi_{a_{cl}},\Delta_{k_{cl}},\Delta^*_{k'_{cl}}$,
one should note that these quantities are the components of the global minimum point of the TDP.

It is easy to understant that due to the invariance of the Lagrangian (\ref{CSC}) and path integration measure with respect to the Abelian $U_{I_3}(1)$ and $U_{AI_3}(1)$ symmetries (\ref{2}), the total TDP $\Omega_{CSC}$ vs $\sigma_{cl}$ and $\pi_{a_{cl}}$ 
exists as a function of two combinations of these quantities, $\Sigma$ and $\Pi$, which are defined around Eq. (\ref{8}) \footnote{A similar procedure was 
carried out in Section \ref{aaa1} in the framework of the simpler NJL model (\ref{1}), and it can be easily repeated in a more general model (\ref{100}).}.
To discuss the symmetry properties of the TDP $\Omega_{CSC}$ with respect to the color $SU_c(3)$ group and its dependence on scalar color diquarks $\Delta_{k_{cl}},\Delta^*_{k'_{cl}}$, it is necessary to take into account the fact that while the quantized fields from the auxiliary Lagrangian (\ref{CSC}) are transformed according to the formulas 
$\psi\to U\psi$, $\Delta\to U^*\Delta$ and $\Delta^*\to U\Delta^*$ (where $U\in SU_c(3)$),
the transformations of the classical diquark fields $\Delta_{cl}$ and $\Delta^*_{cl}$, included in the set of arguments of the thermodynamic potential $\Omega_{CSC}$, have the following form: $\Delta_{cl}\rightarrow \Delta'_{cl} = (U^*)^{-1}\Delta_{cl}=U^T\Delta_{cl},\;\Delta^*_{cl}\rightarrow \Delta^{\prime *}_{cl}=  U^{-1}\Delta^*_{cl}$ (for details, see in Appendix A). In terms of scalar diquark field color components it means that 
\begin{eqnarray}
\Delta'_{a_{cl}} = \Delta_{b_{cl}}U_{ba},~~~~\Delta^{\prime *}_{a_{cl}}=  \left(U^{-1}\right)_{ab}\Delta^*_{b_{cl}}.
\label{102}
\end{eqnarray}
Then, due to the invariance both of the $L_{CSC}$ (\ref{100})-(\ref{CSC}) and path integration measures ${\cal D}\overline\psi 
{\cal D}\psi$, $\prod_{i=1}^3{\cal D}\Delta_i{\cal D}\Delta^*_i$ and $ {\cal D}\sigma \prod_{i=1}^3{\cal D}\pi_i$ in its partition functional under $SU_c(3)$ group, it can be shown that effective action $\Gamma _{CSC}(\sigma_{cl},\pi_{a_{cl}}, \Delta_{cl},\Delta^{*}_{cl})$ is invariant with respect to $SU_c(3)$. As a result, one can conclude that the TDP $\Omega_{CSC}$ of the model is also invariant under $SU_c(3)$ group, i.e. 
$$\Omega_{CSC}(\sigma_{cl},\pi_{a_{cl}}, \Delta_{cl}',\Delta^{\prime *}_{cl})=\Omega_{CSC}(\sigma_{cl},\pi_{a_{cl}}, \Delta_{cl},\Delta^{*}_{cl}).$$
So, instead of six color arguments, the TDP $\Omega_{CSC}$ has to depend only on the single color field combination, $\Delta\equiv \left(\Delta_{1_{cl}}\Delta^*_{1_{cl}} +\Delta_{2_{cl}}\Delta^*_{2_{cl}}+\Delta_{3_{cl}}\Delta^*_{3_{cl}}\right)^{1/2}$. As it is clear from Eq. (\ref{102}), the quantity $\Delta$ is invariant under $SU_c(3)$ transformations.
(Some details of this conclusion can be clarified on the basis of the material presented in Appendix A.) 

Hence the total TDP $\Omega_{CSC}$ of the NJL model (\ref{100})-(\ref{CSC}) depends only on three field combinations, i.e. 
\begin{eqnarray}
\Omega_{CSC}(\sigma_{cl},\pi_{a_{cl}}, \Delta_{k_{cl}},\Delta^{*}_{k'_{cl}})\equiv\Omega_{CSC}(\Sigma,\Pi,\Delta).\label{103}
\end{eqnarray}
So if in the global minimum point of the TDP $\Omega_{CSC}(\Sigma,\Pi,\Delta)$ the quantity $\Delta$ is not zero, then at 
least one of the quantities $\Delta_{k_{cl}}$ is also not equals zero, and color $SU_c(3)$ symmetry is spontaneously broken. In this case the diquark pairing occurs, and the so-called color 
superconducting phase is realized in the NJL model (\ref{100}), etc. 

Recall that in the previous section III we have proved that the Lagrangian (\ref{1}) of the simplest massless NJL model is invariant under the dual transformation ${\cal D}$ (\ref{20}) which leads to the duality between CSB and charged PC phases. Now let us prove that the same property is inherent in the NJL model (\ref{100}) with additional diquark interaction channel.  

To verify that Lagrangian $L_{CSC}$ is invariant under the dual transformation ${\cal D}$, it is necessary to check on this invariance only the second term on the right side of the equality (\ref{100}). But one can show this by the following simple transformations (see also Eqs. (\ref{10})-(\ref{12})),
\begin{eqnarray}
\Delta_k(x)\sim\overline{\psi^{c}} \gamma^{5}\tau_{2}\epsilon_k \psi&=&\psi^{T}C \gamma^{5}\tau_{2}\epsilon_k \psi=
\psi_{R}^{T}C \tau_{2}\epsilon_k \psi_{R}
-\psi_{L}^{T}C \tau_{2}\epsilon_k\psi_{L}\nonumber\\
&\stackrel{{\cal D}}{\longrightarrow}& \psi_{R}^{T}C \tau_{2}\epsilon_k \psi_{R}-\psi_{L}^{T}i\tau_{1}C
\tau_{2}\epsilon_k i\tau_{1}\psi_{L}\nonumber\\
&=&\psi_{R}^{T}C \tau_{2}\epsilon_k \psi_{R}-\psi_{L}^{T}C\tau_{2}\epsilon_k\psi_{L}=\overline{\psi^{c}}
\gamma^{5}\tau_{2}\epsilon_k \psi. \label{300}
\end{eqnarray}
In addition, it is also easy to establish the validity of a similar relation,
\begin{eqnarray}
\Delta_k^*(x)\sim\overline\psi \gamma^{5} \tau_{2}\epsilon_k\psi^c
&\stackrel{{\cal D}}{\longrightarrow}\overline\psi \gamma^{5} \tau_{2}\epsilon_k\psi^c.\label{310}
\end{eqnarray}
Hence, $L_{CSC}$ as a whole is invariant under the  duality transformation ${\cal D}$ (\ref{20}). The auxiliary Lagrangian $\widetilde L_{CSC}$ (\ref{CSC}) of the model is also invariant with respect to the dual transformation, if we add to the corresponding transformations of the spinor fields (\ref{20}) the transformations (\ref{21}) of the auxiliary meson fields $\sigma(x)$ and $\vec\pi_a(x)$, and also take into account that the auxiliary diquark fields do not change (see in Eqs. (\ref{300})-(\ref{310})).

Now, it is important to note that in Appendix C and D the invariance of the measure ${\cal D}\overline\psi{\cal D}\psi$ of the generating functional (\ref{101}) under the duality transformation (\ref{20}) is proved (invariance of the measure ${\cal D}\sigma\prod_{i=1}^3{\cal D}\pi_i\,\prod_{k=1}^3{\cal D}\Delta_k{\cal D}\Delta_k^*$ under PG transformations (\ref{21}) and unchanged scalar diquark fields $\Delta_k$ and $\Delta_k^*$ is rather evident, as it is discussed in Appendix C). In Appendix D it is considered in detail if regularization procedure, which is inevitable in effective models, can break duality of the model. As a consequence (see Appendix C), not only the auxiliary Lagrangian $\widetilde L_{CSC}$, but also the full effective action $\Gamma_{CSC}(\sigma_{cl},\vec\pi_{cl},\Delta_{k_{cl}},\Delta^*_{k'_{cl}})$ and the full TDP $\Omega_{CSC} (\Sigma,\Pi,\Delta)$ (\ref{103}) of the 
model are invariant  
under the PG transformation (\ref{C10})-(\ref{C11}) of the classical fields $\sigma_{cl}(x)$ and $\pi_{a_{cl}}(x)$ (and unchanged classical diquark fields $\Delta_{k_{cl}}$ and $\Delta^*_{k'_{cl}}$), whereas $\mu_I\leftrightarrow\mu_{I5}$. As a whole, we call this procedure the dual transformation $\cal D$ of the TDP. But the last property, i.e. the ${\cal D}$ symmetry of the full TDP, 
is nothing else than its invariance under the dual transformation $\widetilde{\cal D}$ 
(\ref{9}) (when 
$\Sigma\leftrightarrow\Pi$,  $\mu_I\leftrightarrow\mu_{I5}$ and $\Delta$ is unchanged) of the TDP found in the mean-field approximation \cite{u2}. So the $\cal D$ symmetry of the TDP should manifest itself in the full phase portrait of the model in the form of duality between the CSB and charged PC phenomena.

As a result, we see that the dual symmetry between CSB and charged PC phenomena of dense quark medium with 
isospin and chiral 
isospin asymmetries (see the description below Eq. (\ref{9})), predicted earlier in different NJL models \cite{kkz18-2,u2} within the 
framework of the mean-field  approximation to the phase structure, is in fact a genuine property of dense quark matter described by 
the massless NJL models (\ref{1}) and (\ref{100}). So the duality between CSB and charged PC phenomena is realized in the full 
$(\mu_B,\mu_I,\mu_{I5},\mu_5)$-phase diagrams of these models (and not only in the scope of the mean-field approximation), 
since it is a consequence of the dual symmetry ${\cal D}$ (\ref{20}) both of the microscopic
NJL Lagrangians (\ref{1}) and (\ref{100}) and the measure ${\cal D}\overline\psi{\cal D}\psi$ and ${\cal D}\sigma\prod_{i=1}^3{\cal D}\pi_i\,\prod_{k=1}^3{\cal D}\Delta_k{\cal D}\Delta_k^*$ of path integration.

\section{Dual symmetry of dense QCD }
In the previous Sections~III and~IV, we established an important result: two formally different massless NJL models, (\ref{1}) and (\ref{100}), which nevertheless effectively describe the low-energy and moderate-density regimes of massless QCD, both possess the dual symmetry ${\cal D}$~(\ref{20}) between the CSB and charged PC phenomena.  
This naturally suggests that the Lagrangian of massless dense QCD itself should also be invariant under the same dual transformation~${\cal D}$.  
In this section we demonstrate that this is indeed the case and discuss several related properties of QCD, which confirm that a dual symmetry between the CSB and charged PC phases must also exist in fundamental QCD.

As a starting point, we consider the Lagrangian for the quark sector of massless QCD, extended by the baryon, isospin, chiral, and chiral isospin chemical potentials:
\begin{equation}
L_{QCD}=i\overline\psi \gamma^{\nu}\nabla_{\nu}\psi+\frac{\mu_B}{3}\overline{\psi}\gamma^{0}\psi
+\mu_5\,\overline{\psi}\gamma^{0}\gamma^5\psi
+\frac{\mu_I}{2}\,\overline{\psi}\gamma^{0}\tau_{3}\psi
+\frac{\mu_{I5}}{2}\,\overline{\psi}\gamma^{0}\gamma^{5}\tau_{3}\psi,
\label{46}
\end{equation}
where $\psi(x)$ is the two-flavor quark doublet.  
The covariant derivative $\nabla_{\nu}=\partial_\nu-ig\lambda_aA^a_\nu(x)$ acts in color space, with $\lambda_a$ ($a=1,\ldots,8$) being the $3\times3$ Gell--Mann matrices and $A^a_\nu(x)$ the $SU_c(3)$ gauge fields.  
The Lagrangian~(\ref{46}) therefore describes dense quark matter with both isospin and chiral asymmetries.

We now show that the dual symmetry ${\cal D}$, defined in Eq.~(\ref{20})---which in the NJL framework relates the CSB and charged PC channels---is also realized in the fundamental QCD Lagrangian~(\ref{46}).  
Indeed, it follows from the discussion in the previous sections that the sum of all chemical-potential terms in Eq.~(\ref{46}) is invariant under ${\cal D}$.  
The kinetic term is also invariant, since
\begin{eqnarray}
i\overline\psi \gamma^{\nu}\nabla_{\nu}\psi
&=&i\overline\psi_{R}\gamma^{\nu}\nabla_{\nu}\psi_{R}
+i\overline\psi_{L}\gamma^{\nu}\nabla_{\nu}\psi_{L}
\stackrel{{\cal D}}{\longrightarrow}
i\overline\psi_{R}\gamma^{\nu}\nabla_{\nu}\psi_{R}
+i\overline\psi_{L}(-i\tau_{1})\gamma^{\nu}\nabla_{\nu}i\tau_{1}\psi_{L}
\nonumber\\[4pt]
&=&i\overline\psi_{R}\gamma^{\nu}\nabla_{\nu}\psi_{R}
+i\overline\psi_{L}\gamma^{\nu}\nabla_{\nu}\psi_{L}
=i\overline\psi\gamma^{\nu}\nabla_{\nu}\psi,
\label{47}
\end{eqnarray}
and thus the full QCD Lagrangian~(\ref{46}) is symmetric under the dual transformation ${\cal D}$~(\ref{20}).

It is well known from quantum field theory that the invariance of a classical Lagrangian under a certain group of transformations does not necessarily imply the invariance of the corresponding quantum theory under the same transformations (see, for example, Section~9.6 of Ref.~\cite{Peskin}).  
Such discrepancies---known as anomalies---arise when the symmetry of the classical action is not preserved in the quantized theory.  
This typically occurs for axial transformations of spinor fields.  
To verify that a given symmetry group~$G$ is non-anomalous, one must explicitly demonstrate the invariance of the functional integration measure ${\cal D}\overline\psi{\cal D}\psi$ in the model's partition function under transformations of~$G$.

For illustration, we refer the reader to Appendix~A, where the invariance of the $U_{I_3}(1)$ symmetry~(\ref{50}) is demonstrated for both the auxiliary NJL Lagrangian $\widetilde L_{NJL}$ of Eq.~(\ref{1}) and its effective action $\Gamma(\sigma_{cl},\pi_{a\,cl})$.  
There it is shown that the key step in establishing this invariance is proving that the integration measure is invariant under the $U_{I_3}(1)$ transformation.

As shown above, the QCD Lagrangian~(\ref{46}) is symmetric with respect to the dual transformation ${\cal D}$~(\ref{20}).  
Since this transformation involves the spinor matrix $\gamma^5$, one might suspect that ${\cal D}$ could be anomalous.  
If that were the case, neither the full effective action nor the thermodynamic potential would remain dually symmetric, and the dual correspondence between the CSB and charged PC phases of dense quark matter would be lost.  
However, as proven in Appendix~C, this suspicion does not hold: the functional integration measure in the QCD generating functional is in fact invariant under the dual transformation ${\cal D}$.

Specifically, Appendix~C shows that the measure ${\cal D}\overline\psi{\cal D}\psi$ is invariant under both vector transformations $U(\vec\theta)\in SU_V(2)$ and axial transformations $U_A(\vec\omega)\in SU_A(2)$ (see Eqs.~(\ref{C1}) and~(\ref{C2}), respectively).  
Consequently, it is invariant under their product $U(\vec\theta)U_A(\vec\omega)$.  
By choosing the specific parameters $\vec\theta=-\vec\omega=(\pi/4,0,0)$, this combined transformation coincides exactly with the dual (Pauli--Gursey) transformation ${\cal D}$~(\ref{20}) of the spinor doublet~$\psi(x)$.  
Therefore, the path-integration measure ${\cal D}\overline\psi{\cal D}\psi$ in massless QCD is invariant under ${\cal D}$, and this transformation is \emph{non-anomalous}.  

It follows that both the effective action and the thermodynamic potential of QCD are symmetric under ${\cal D}$.  
Hence, in the full $(\mu_B,\mu_I,\mu_{I5},\mu_5)$ phase diagram, the CSB phase is dually conjugated to the charged PC phase.  
This dual symmetry should be regarded as a fundamental property of massless QCD, expected to manifest itself in any consistent approximation to the QCD phase structure.

\section{Summary and conclusions}

In this work, we have demonstrated that the Lagrangians of the massless NJL models~(\ref{1}) and~(\ref{100}), constructed from three-color $u$ and $d$ quarks and extended by the baryon, isospin, chiral, and chiral isospin chemical potentials $\mu_B$, $\mu_I$, $\mu_{I5}$, and $\mu_5$, possess a discrete dual symmetry ${\cal D}$~(\ref{20}) relating the chiral symmetry breaking (CSB) and charged pion condensation (PC) channels of interaction (see Sections~II--IV).  
Since the path integration measure ${\cal D}\overline\psi{\cal D}\psi$ in the generating functionals of these models is also invariant under the ${\cal D}$ transformation (as shown in Appendix~C), it follows that the corresponding total thermodynamic potentials, viewed as functions of the chemical potentials and the order parameters $\Sigma$ and $\Pi$ associated with the CSB and charged PC phases, are symmetric under the dual transformation~(\ref{9}).  

As a consequence, in the \textit{complete} phase diagram of each model, the CSB and charged PC phases are arranged as dual, mirror-symmetric counterparts of one another.  
In particular, at fixed $\mu_B$ and $\mu_5$, the $(\mu_I,\mu_{I5})$ phase portrait exhibits mirror symmetry with respect to the line $\mu_I=\mu_{I5}$.  
Thus, we have shown that the dual symmetry between the CSB and charged PC phenomena is an intrinsic property of the full massless NJL models~(\ref{1}) and~(\ref{100}), not merely of their mean-field (leading $1/N_c$) approximation as previously established in Refs.~\cite{kkz18,kkz18-2,u2}.  

Furthermore, in Section~V, we proved that the same dual symmetry ${\cal D}$~(\ref{20}) is also realized in the fundamental massless QCD Lagrangian~(\ref{46}) for three-colored $u$ and $d$ quarks.  
This result implies that the dual correspondence between CSB and charged PC phenomena is not an artifact of effective theories but a \emph{fundamental property of dense massless QCD}.  
Hence, it should manifest itself within any consistent approximation to the QCD phase diagram.  
Moreover, in the case of two-color QCD, additional dual correspondences appear, enriching the symmetry structure of the phase diagram~\cite{Khunjua:2024kdc_a,Khunjua:2024kdc_b}.  

Establishing this dual correspondence directly from first principles, i.e. within QCD itself, opens the possibility of a broad class of applications of dual symmetries in the study of the QCD phase structure.  
The concept of duality is no longer confined to the chiral dynamics of quarks: it can, in principle, encompass other QCD phenomena, including gluonic degrees of freedom, transitions between confined and deconfined phases, and topological properties.  
Moreover, the dual symmetry is not limited to the validity range of effective models; it should hold at any energy scale and for both real and imaginary values of the chemical potentials, potentially offering new insights into the analytic continuation of QCD thermodynamics.  

Finally, we note that the dual symmetry ${\cal D}$ has been rigorously established here for massless quarks.  
In realistic situations, where quarks are massive, remnants of this symmetry should persist to some degree. A physical current mass breaks it (under the Pauli--Gursey rotation~(\ref{10}) the mass term $m_0\overline\psi\psi$ turns into the charged-pion channel Eqs.~(\ref{13})--(\ref{14})), though the breaking is mild, it is set by the tiny mass $m_0\simeq5.5$~MeV compared to far larger scales of dense matter. 
Indeed, the mean-field analysis of NJL model in the physical point (with non-zero current masses)~\cite{kkz19} demonstrates that the CSB--PC duality becomes approximate but remains clearly visible and is an excellent approximation at chemical potentials larger than half of pion mass. (Indeed,
once chemical potentials climb above some very reasonable value CSB and charged PC regions look practically as in the chiral limit.) 
So nothing points to why one could not expect that it is good approximation in general, not only in mean-field. 
For simplicity, the present study has been restricted to zero temperature ($T=0$), but we see no compelling reason why the duality ${\cal D}$ in the phase structure of dense quark matter should disappear at finite temperature.  
We therefore expect the ${\cal D}$-duality to remain a robust and general feature of QCD matter under a wide range of conditions.

\begin{acknowledgments}
The work of T.~G.~Khunjua was supported by Shota Rustaveli National Science Foundation of Georgia (SRNSFG) [grant number FR-25-7544, ``Nonperturbative Analysis of the QCD Critical Point: Regularization problems in the Effective Models and Renormalization Group Techniques''].
\end{acknowledgments}

\appendix

\section{Symmetry properties of effective action and TDP}\label{ApA}

$\bullet$ First of all, here we consider the symmetry properties of the effective action 
\begin{eqnarray}
\Gamma_{\cal M} (\sigma_{cl},\pi_{a_{cl}})\equiv\Gamma_{\cal M} (\sigma_{cl},\pi_{3_{cl}};\pi_{1_{cl}},\pi_{2_{cl}})={\cal W}_{\cal M}(J)-\int d^4x\left\{\sigma_{cl}(x)J_0(x)+\sum_{a=1}^3
\pi_{a_{cl}}(x)J_a(x)\right\},\label{aa6}
\end{eqnarray}
where 
\begin{eqnarray}
\sigma_{cl}(x)\equiv\frac{\delta {\cal W}_{\cal M}(J)}{\delta J_0(x)}=\frac{1}{{\cal Z}_{\cal M}(J)}
\int {\cal D}\overline\psi 
{\cal D}\psi {\cal D}\sigma \prod_{a=1}^3{\cal D}\pi_a\sigma (x)\exp\left\{i\int d^4y\big [ 
\widetilde L_{NJL}+N_c(\sigma J_0+\pi_a J_a)\big]\right\}=\langle\sigma (x)\rangle_J,\nonumber\\
\pi_{k_{cl}}(x)\equiv\frac{\delta {\cal W}_{\cal M}(J)}{\delta J_k(x)}=\frac{1}{{\cal Z}_{\cal M}(J)}
\int {\cal D}\overline\psi 
{\cal D}\psi {\cal D}\sigma \prod_{a=1}^3{\cal D}\pi_a\pi_k (x)\exp\left\{i\int d^4y\big [ 
\widetilde L_{NJL}+N_c(\sigma J_0+\pi_a J_a)\big]\right\}=\langle\pi_k (x)\rangle_J,
\label{aa1}\end{eqnarray}
which follow from the invariance of the auxiliary Lagrangian $\widetilde L_{NJL}$ (\ref{3}) under 
the transformations (\ref{2}), (\ref{50}) and (\ref{5}).

Now, let us study how the invariance of the auxiliary Lagrangian (\ref{3}) under these transformations 
is realized in terms of the functional ${\cal W}_{\cal M}(J)$ (\ref{a3}).
For simplicity, we consider in detail only the role of the subgroup $U_{I_3}(1)$. In this connection, we replace the integration variables in this continual integral 
according to the rule: $\psi(x)\to\psi' (x)=\exp (i\alpha\tau_3/2)\psi (x)$, 
$\left( {\begin{array}{c}
\pi_1 \\
\pi_2
\end{array} } \right)\to \left( {\begin{array}{c}
\pi_1' \\
\pi_2'
\end{array} } \right)= O(\alpha)\left( {\begin{array}{c}
\pi_1 \\
\pi_2
\end{array} } \right)$, where $O(\alpha)$ is a 2$\times$2 matrix of the form $O(\alpha)=
\left( {\begin{array}{c}
\cos\alpha, \sin\alpha \\
-\sin\alpha, \cos\alpha
\end{array} } \right)$ and scalar fields $\sigma(x),\pi_3(x)$ remain intact. 
This change of variables is nothing more than a transformation of Fermi fields under the action 
of $U_{I_3}(1)$ (\ref{2}), and of scalar fields according to the rule (\ref{50}).
Then it is 
necessary take into account that in this case neither the integration measure, nor the Lagrangian change. In this case, a significant portion of the members with sources, which includes only the $\pi_1$ and $\pi_2$  scalar fields,  is transformed as follows (summation over the repeated indices is implied from 1 to 2):
\begin{eqnarray}
\pi_aJ_a\to\pi'_aJ_a=O(\alpha)_{ai}\pi_iJ_a=\pi_iO^T(\alpha)_{ia}J_a\equiv\pi_iJ'_i,
\label{aa2}
\end{eqnarray} where
$J_i'=O^T(\alpha)_{ia}J_a$, i.e. $J_1'=\cos\alpha J_1-\sin\alpha J_2$ and $J_2'=\sin\alpha J_1+\cos\alpha J_2$.
Hence, it is easy to see that 
\begin{eqnarray}
{\cal W}_{\cal M}(J)\equiv {\cal W}_{\cal M}(J_0,J_3|J_1,J_2)=
{\cal W}_{\cal M}(J_0,J_3|J_1',J_2').
\label{aa3}
\end{eqnarray}
It means that functional 
${\cal W}_{\cal M}(J_0,J_3|J_1,J_2)$ is invariant under the $SO(2)$ group acting on two sources, $J_1$ and $J_2$, and it does not change its meaning when $J_1,J_2\to J_1',J_2'$. Now let us do the Legendre transform of the functional ${\cal W}_{\cal M}(J_0,J_3|J_1',J_2')$:
\begin{eqnarray}
\Gamma_{\cal M} (\sigma_{cl},\pi_{3_{cl}};\pi_{1_{cl}}',\pi_{2_{cl}}')={\cal W}_{\cal M}(J_0,J_3|J_1',J_2')-\int d^4x\left\{\sigma_{cl}(x)J_0(x)+\pi_{3_{cl}}(x)J_3(x)+\sum_{a=1}^2
\pi_{a_{cl}}'(x)J_a'(x)\right\},\label{aa4}~~~~~~~~~
\end{eqnarray}
where 
\begin{eqnarray}\pi_{a_{cl}}'(x)=\frac{\delta {\cal W}_{\cal M}(J_0,J_3|J_1',J_2')}{\delta J_a'(x)}=\int d^4z
\frac{\delta {\cal W}_{\cal M}(J_0,J_3|J_1,J_2)}{\delta J_i(z)}\frac{\delta J_i(z)}{\delta J_a'(x)}=
\pi_{i_{cl}}(x)O(\alpha)_{ia}.
\label{aa5}
\end{eqnarray}
To obtain the last expression in Eq. (\ref{aa5}), we use the relations $J_i(z)=
O(\alpha)_{ia}J_a'(z)$ and $\delta J_i(z)/\delta J_a'(x)=O(\alpha)_{ia}\delta(x-z)$. Moreover, here the equality (\ref{aa3}) for
functional ${\cal W}_{\cal M}(J_0,J_3|J_1',J_2')$ as well as the definition (\ref{aa1}) of 
$\pi_{i_{cl}}(x)$ are also taken into account. It is clear from Eqs. (\ref{aa5}) and (\ref{aa2}) that $\pi_{a_{cl}}'(x)J_a'(x)=\pi_{a_{cl}}(x)J_a(x)$. Now, taking into account all of the above, we see that
\begin{eqnarray}
\Gamma_{\cal M} (\sigma_{cl},\pi_{3_{cl}};\pi_{1_{cl}}',\pi_{2_{cl}}')={\cal W}_{\cal M}(J)-\int d^4x\left\{\sigma_{cl}(x)J_0(x)+\sum_{a=1}^3
\pi_{a_{cl}}(x)J_a(x)\right\}\equiv\Gamma_{\cal M} (\sigma_{cl},\pi_{3_{cl}};\pi_{1_{cl}},\pi_{2_{cl}}),\label{aa7}
\end{eqnarray}
i.e. effective action (\ref{aa6}) or (\ref{a6}) remains invariant when $\pi_{1_{cl}}(x)$ and $\pi_{2_{cl}}(x)$
transform as $SO(2)$ doublet. In a similar way it is possible to show that effective action 
$\Gamma (\sigma_{cl},\pi_{3_{cl}};\pi_{1_{cl}},\pi_{2_{cl}})$ is also invariant when 
$\sigma_{cl}(x)$ and $\pi_{3_{cl}}(x)$ transform as $SO(2)$ doublet. Hence 
\begin{eqnarray}
\Gamma_{\cal M} (\sigma_{cl}',\pi_{3_{cl}}';\pi_{1_{cl}}',\pi_{2_{cl}}')=
\Gamma_{\cal M} (\sigma_{cl},\pi_{3_{cl}};\pi_{1_{cl}},\pi_{2_{cl}}),\label{aa8}
\end{eqnarray}
where 
$\left( {\begin{array}{c}
\pi_{1_{cl}}' \\ \pi_{2_{cl}}'\end{array} } \right)= \left( {\begin{array}{c}
\cos\alpha, -\sin\alpha \\
\sin\alpha, \cos\alpha
\end{array} } \right)\left( {\begin{array}{c}
\pi_{1_{cl}} \\ \pi_{2_{cl}} \end{array} } \right)$ and $\left( {\begin{array}{c}
\sigma_{cl}' \\ \pi_{3_{cl}}'\end{array} } \right)= \left( {\begin{array}{c}
\cos\beta, -\sin\beta \\
\sin\beta, \cos\beta
\end{array} } \right)\left( {\begin{array}{c}
\sigma_{cl} \\ \pi_{3_{cl}} \end{array} } \right)$. Then, if scalar fields $\sigma_{cl}$ and $\pi_{a_{cl}}$
do not depend on spacetime coordinates, it is easy to conclude that thermodynamic potential 
$\Omega_{\cal M} (\sigma_{cl},\pi_{a_{cl}})$ depends effectively on two variables, $\Sigma=\sqrt{\sigma_{cl}^2+
\pi_{3_{cl}}^2}$ and $\Pi=\sqrt{\pi_{1_{cl}}^2+\pi_{2_{cl}}^2}$.

$\bullet$ Now, in the same way, let us discuss the transformation properties of the effective action 
$\Gamma_{CSC}(\sigma_{cl},\pi_{a_{cl}}, \Delta_{k_{cl}},\Delta^*_{k'_{cl}})$ of the NJL model 
(\ref{100})-(\ref{CSC}) (it is defined in the section IV) with respect to color $SU_c(3)$ group. 
\footnote{Its propeties with respect to the transformations (\ref{2}), (\ref{50}) and (\ref{5}) are 
the same as those of the effective action (\ref{aa6}) of the NJL model (\ref{1}).}
To beging with, let us recall that  auxiliary scalar diquark fields $\Delta_a(x)$ and $\Delta^*_b(x)$ from $\widetilde L_{CSC}$ transform under color $SU_c(3)$ group in the following way 
\begin{eqnarray}
\Delta_a \rightarrow \Delta'_a = U_{ab}^* \Delta_b,\;\;\;\;\;\;\;\;
\Delta_a^*\rightarrow \Delta'^*_a= U_{ab} \Delta_b^*,\label{a99}
\end{eqnarray}
where $U\in SU_c(3)$ is a 3$\times$3 matrix. This fact can be easily established using the equations of motion (\ref{99}) for 
auxiliary diquark fields, in which it is necessary to perform the $SU_c(3)$ transformation of the 
spinor fields, $\psi (x)\to U\,\psi (x)$. It is clear from Eq. (\ref{a99}) that diquark fields $\Delta_a(x)$
form an antitriplet, whereas the set of $\Delta^*_b(x)$ ($a,b=1,2,3$) is a triplet of the color 
$SU_c(3)$ group. Moreover, it is rather evident that the integration measure ${\cal D}\overline\psi 
{\cal D}\psi\prod_{i=1}^3{\cal D}\Delta_i{\cal D}\Delta^*_i$ in the generating functional 
${\cal Z}_{CSC}(J, \overline{\eta}, \eta )$ (\ref{101}) is invariant under $SU_c(3)$ transformations.

Now, let us find how the scalar diquark sources $\overline{\eta}_a$ and $\eta_b$, included in the 
generating functional (\ref{101}), should be transformed so that it remains invariant under 
$SU_c(3)$ transformations. In this case, taking into account that both $\widetilde L_{CSC}$
and the integration measure of this functional is $SU_c(3)$ invariant, we can change variables in 
the continual integral, $\Delta_a \rightarrow \Delta'_a$ and $\Delta_a^*\rightarrow \Delta'^*_a$ as 
in Eq. (\ref{a99}), and then define the $SU_c(3)$ 
transformed sources by the following relations (summation over repeated indices is implied)
\begin{eqnarray}
\overline{\eta}_a \Delta'_a =\overline{\eta}_aU^*_{ab} \Delta_b=\overline{\eta}'_b \Delta_b\Longrightarrow
\overline{\eta}'_b= U_{ba}^\dagger\overline{\eta}_a,\nonumber\\
\eta_a \Delta'^*_a =\eta_aU_{ab}\Delta^*_b=\eta'_b \Delta^*_b\Longrightarrow
\eta'_b= U^T_{ba}\eta_a.\label{a100}
\end{eqnarray}
As a consequence, we can conclude that ${\cal Z}_{CSC}(J, \overline{\eta}, \eta )=
{\cal Z}_{CSC}(J, \overline{\eta}', \eta' )$ and ${\cal W}_{CSC}(J, \overline{\eta}, \eta )=
{\cal W}_{CSC}(J, \overline{\eta}', \eta' )$ (this functional is defined after Eq. (\ref{101})), 
where primed sources are presented in Eq. (\ref{a100}). By means of the second of these invariance relations it is possible to find how classical diquark fields $\Delta_{a_{cl}}(x)$ and $\Delta^*_{b_{cl}}(x)$ transform under $SU_c(3)$. Indeed, 
\begin{eqnarray}
\Delta'_{a_{cl}} &=& \frac{\delta {\cal W}_{CSC}(J,\overline{\eta}', \eta' )}{\delta \overline{\eta}'_{a}(x)} = \int dz \frac{\delta {\cal W}_{CSC}(J,\overline{\eta}, \eta )}{\delta \overline{\eta}_b(z)} \frac{\delta \overline{\eta}_b(z)}{\delta \overline{\eta}'_{a}(x)}=\Delta_{b_{cl}}U_{ba}=(U^*)^{-1}_{ab}\Delta_{b_{cl}},\nonumber\\
\Delta'^*_{a_{cl}} &=& \frac{\delta {\cal W}_{CSC}(J,\overline{\eta}', \eta' )}{\delta \eta'_{a}(x)}
 = \int dz \frac{\delta {\cal W}_{CSC}(J,\overline{\eta}, \eta )}{\delta \eta_b(z)} \frac{\delta \eta_b(z)}{\delta \eta'_{a}(x)}=\Delta^*_{b_{cl}}U^{*}_{ba}=U^{\dagger}_{ab}\Delta^*_{b_{cl}}.\label{a101}
\end{eqnarray}
where we also used the relations (\ref{a100}) and their consequences, $\frac{\delta \overline{\eta}_b(z)}{\delta \overline{\eta}'_{a}(x)}=U_{ba}\delta(z-x)$ and $\frac{\delta \eta_b(z)}{\delta \eta'_{a}(x)}=U^*_{ba}\delta(z-x)$. 

Now, using the transformations (\ref{a100}) of the sources as well as the transformations (\ref{a101}) of classical diquark fields under $SU_c(3)$, it is easy to establish that the quantity $(\overline{\eta_a}\, \Delta_{a_{cl}}+\eta_b \Delta^*_{b_{cl}})$, which is present in the definition of the full effective action $\Gamma _{CSC}(\sigma_{cl},\pi_{a_{cl}}, \Delta_{k_{cl}},\Delta^{*}_{l_{cl}})$ of the NJL model (\ref{100})-(\ref{CSC}) (see in the section IV), remains intact under simultaneous transformations (\ref{a100})-(\ref{a101}). Finally, if we now recall that the functional ${\cal W}_{CSC}(J, \overline{\eta}, \eta )$ is also $SU_c(3)$ invariant, then it is easy to conclude that the full effective action $\Gamma _{CSC}(\sigma_{cl},\pi_{a_{cl}}, \Delta_{k_{cl}},\Delta^{*}_{l_{cl}})$ of the NJL model (\ref{100}) (it is introduced in the section IV) is invariant with respect to color transformations (\ref{a101}) of classical scalar diquark fields, i.e.
$$\Gamma _{CSC}(\sigma_{cl},\pi_{a_{cl}}, \Delta'_{k_{cl}},\Delta'^{*}_{l_{cl}})=\Gamma _{CSC}(\sigma_{cl},\pi_{a_{cl}}, \Delta_{k_{cl}},\Delta^{*}_{l_{cl}}).$$
As a consequence, the total TDP $\Omega_{CSC}(\sigma_{cl},\pi_{a_{cl}}, \Delta'_{k_{cl}},\Delta'^{*}_{l_{cl}})$ (\ref{TDP}) of the model is also $SU_c(3)$ invariant, so it has to depend 
only on the combination $\Delta\equiv\sqrt{\Delta_{a_{cl}}\Delta^*_{a_{cl}}}$, which is invariant under $SU_c(3)$ as it can be easily checked using Eqs. (\ref{a101}). 

\section{Effective action of the NJL model (\ref{1}) in the leading large-$N_c$ order}\label{ApB}

Starting from the partition function (\ref{a9}), it is convenient to use the steepest descent method 
(see, e.g., Ref. \cite{iliopoulos}) when calculating the effective action (\ref{a6}) or (\ref{aa6})
of the simplest NJL model (\ref{1}). To begin with, suppose that the set of quantities $\sigma_0\equiv\sigma_0(x;J)$ and $\vec\pi_{0}\equiv\vec\pi_{0}(x;J)$ is the solution of the following system of four equations:
\begin{eqnarray}
\frac{\delta S_{eff}(\sigma,\vec\pi)}{\delta \sigma (x)}+J_0(x)=0,~~~
\frac{\delta S_{eff}(\sigma,\vec\pi)}{\delta \pi_k(x)}+J_k(x)=0,~~~~k=1,2,3, \label{b5}
\end{eqnarray}
where $S_{eff}(\sigma,\vec\pi)$ is defined in Eq. (\ref{a10}). Another name for the set $(\sigma_0,\vec\pi_0)$ is the saddle point of the functional $S_{eff}(\sigma,\vec\pi)$. (Note that the quantities $\sigma_0$ and $\pi_{0k}$
do not depend on $N_c$, i.e. they have the order of magnitude $N_c^0$ for large $N_c$.) Then, expanding the functional $S_{eff}(\sigma,\vec\pi)$  in a Taylor series around its saddle point $(\sigma_0,\vec\pi_0)$,
\begin{eqnarray}
S_{eff}(\sigma,\vec\pi)=S_{eff}(\sigma_0,\vec\pi_{0})
+\int d^4x \frac{\delta S_{eff}}{\delta\sigma (x)}\Big |_{(\sigma_0,\vec\pi_0)}\tilde\sigma 
(x)+\int d^4x \frac{\delta S_{eff}}{\delta\pi_k (x)}\Big |_{(\sigma_0,\vec\pi_0)}\tilde\pi_k(x)+\Delta S_{eff}(\tilde\sigma (x),\tilde\pi_k (x)).\label{b6}
\end{eqnarray}
(The Taylor series expansion of the effective action in the simplest model of field theory 
with one scalar field is presented, for example, in Ref. \cite{Peskin} by formula (11.58).)
Here $\tilde\sigma (x)=\sigma (x)-\sigma_0(x;J)$, $\tilde\pi_k (x)=\pi_k (x)-\pi_{k_0}(x;J)$ 
and by $\Delta S_{eff}(\tilde\sigma (x),\tilde\pi_k (x))$ we denote the sum of all higher order contributions over the quantities $\tilde\sigma (x)$ and $\tilde\pi_k (x)$. Taking into account the saddle point equations (\ref{b5}), we obtain from Eq. (\ref{b6}) 
\begin{eqnarray}
S_{eff}(\sigma,\vec\pi)+\int d^4x\big [ \sigma(x) J_0(x)+\vec\pi(x)\cdot\vec J(x)\big]=
\nonumber~~~~~~~~~~~~~~~~~\\S_{eff}(\sigma_0,\vec\pi_{0})+
\int d^4x\big [ \sigma_0(x) J_0(x)+\vec\pi_0(x)\cdot\vec J(x)\big]
+\Delta S_{eff}(\tilde\sigma (x),\tilde\pi_k (x)) \label{b7}
\end{eqnarray}
Substituting this expression into the partition function (\ref{a9}) and changing variables in the path integral, $\sigma\to\tilde\sigma (x)$, $\pi_k\to\tilde\pi_k (x)$ , we have
\begin{eqnarray}
\exp(iN_c{\cal W}_{\cal M}(J))=N'\exp\big(i N_c\big\{S_{eff}(\sigma_0,\vec\pi_{0})+
\int d^4x\big [ \sigma_0(x) J_0(x)+\vec\pi_0(x)\cdot\vec J(x)\big] \big\}\big)\times
\nonumber~~~~~\\
\int {\cal D}\tilde\sigma\prod_{a=1}^3{\cal D}\tilde\pi_a\exp\left (iN_c \Big\{
\Delta S_{eff}(\tilde\sigma (x),\tilde\pi_k (x))\Big\}\right ). \label{b8}
\end{eqnarray}
Now, let us again change variables, this time in the path integral (\ref{b8}), $\tilde\sigma\to
\tilde\sigma/\sqrt{N_c}$ and $\tilde\pi_k\to\tilde\pi_k (x)/\sqrt{N_c}$. Then it takes the form
\begin{eqnarray}
\int {\cal D}\tilde\sigma\prod_{a=1}^3{\cal D}\tilde\pi_a\exp\left (iN_c \Big\{
\Delta S_{eff}(\tilde\sigma (x),\tilde\pi_k (x))\Big\}\right )=
\frac{1}{N_c^2}\int {\cal D}\tilde\sigma\prod_{a=1}^3{\cal D}\tilde\pi_a\exp\left (i \Big\{
A(\tilde\sigma (x),\tilde\pi_k (x))+{\cal O}\left (\frac{1}{\sqrt{N_c}}\right )\Big\}\right ), \label{b9}
\end{eqnarray}
where $A(\tilde\sigma (x),\tilde\pi_k (x))$ is the quadratic part of the functional 
$\Delta S_{eff}(\tilde\sigma (x),\tilde\pi_k (x))$ in $\tilde\sigma (x)$ and $\tilde\pi_k (x)$, i.e.
\begin{eqnarray}
A(\tilde\sigma (x),\tilde\pi_k (x))=
\int d^4x d^4y\Big\{\frac 12\frac{\delta^2 S_{eff}}{\delta\sigma (x)\delta\sigma(y)}\Big |_{(\sigma_0,\vec\pi_0)}\tilde\sigma 
(x)\tilde\sigma(y)+~~~~~~~~~~\nonumber\\
\sum_{k=1}^3\frac{\delta^2 S_{eff}}{\delta\sigma (x)\delta\pi_k(y)}\Big |_{(\sigma_0,\vec\pi_0)}\tilde\sigma 
(x)\tilde\pi_k(y)+
\frac 12\sum_{k,l=1}^3\frac{\delta^2 S_{eff}}{\delta\pi_k (x)\delta\pi_l(y)}\Big |_{(\sigma_0,\vec\pi_0)}\tilde\pi_l 
(x)\tilde\pi_k(y)\Big\}. \label{b10}
\end{eqnarray}
Now, taking into account at $N_c\to\infty$ the general relation $\exp \left (A+{\cal O}(1/\sqrt{N_c} )\right )=\exp (A)\big\{1+{\cal O}(1/\sqrt{N_c})\big\}$, one can take the logarithm of both sides of Eq. (\ref{b8}) and obtain (up to irrelevant constants) the functional $W(J)$ in the learding order of the large-$N_c$ technique:
\begin{eqnarray}
{\cal W}_{\cal M}(J)=S_{eff}(\sigma_0,\vec\pi_{0})+
\int d^4x\big [ \sigma_0(x) J_0(x)+\vec\pi_0(x)\cdot\vec J(x)\big]+{\cal O}(1/N_c). \label{b11}
\end{eqnarray}
Recall that the saddle point $(\sigma_0,\vec\pi_0)$ depends on external sources $J_0(x)$ and $J_k(x)$. Now we are going to find the effective action $\Gamma (\sigma_{cl},\vec\pi_{cl})$ (\ref{aa6}) (or (\ref{a6})) starting from Eq. (\ref{b11}). To this end, first of all it is necessary to find the classical fields $\sigma_{cl}(x)\equiv \delta {\cal W}_{\cal M}(J)/\delta J_0(x)$ and $\pi_{k_{cl}}(x)\equiv \delta {\cal W}_{\cal M}(J)/\delta J_k(x)$, and it follows from Eq. (\ref{b11}) that
\begin{eqnarray}
\sigma_{cl}(x)=\sigma_0(x;J)+\int d^4y\left(\frac{\delta S_{eff}}{\delta\sigma_0(y;J)}+J_0(y)\right)\frac{\delta\sigma_0(y;J)}{\delta J_0(x)}+\nonumber\\
\int d^4y\left(\frac{\delta S_{eff}}{\delta\pi_{k_0}(y;J)}+J_{k}(y)\right)\frac{\delta\pi_{k_0}(y;J)}{\delta J_0(x)}+{\cal O}(1/N_c). \label{b12}
\end{eqnarray}
But the expressions in parentheses in Eq. (\ref{b12}) are equal to zero (due to the saddle point equations (\ref{b5})). Hence in the large-$N_c$ limit $\sigma_{cl}(x)=\sigma_0(x;J)$. In a similar way, it is possible to show that in the same approximation $\vec\pi_{cl}(x)=\vec\pi_0(x;J)$. Applying this result in Eq. (\ref{b11}) and in the definition of the effective action $\Gamma (\sigma_{cl},\vec\pi_{cl})$ (\ref{a6}), we see that 
\begin{eqnarray}
\Gamma (\sigma_{cl},\vec\pi_{cl})=S_{eff}(\sigma_{cl},\vec\pi_{cl})+{\cal O}(1/N_c). \label{b13}
\end{eqnarray}
Comparing our result (\ref{b13}) with the expression obtained for the effective action 
(potential) in the mean-field approximation (see, for example, in Ref. \cite{buballa_a,buballa_b,buballa_c,buballa_d,buballa_e,buballa_f}), 
we see that they are identical. Therefore, in what follows we will use the notation 
$\Gamma_{mf} (\sigma_{cl},\vec\pi_{cl})$ for the effective action calculated both in the 
mean-field and in the leading order of the $1/N_c$-approximations (see in Eq. (\ref{a11})). 

\section{Dual invariance of path integration measure} \label{ApC}

Let us show that path integration measure ${\cal D}\overline\psi 
{\cal D}\psi$ in the partition function (\ref{a1}) is invariant under Pauli-Gursey 
(or duality) transformation (\ref{10})-(\ref{12}). Recall that $\psi\equiv\psi(x)$ is a flavor doublet, i.e. $\psi=\left( {\begin{array}{c}
\psi_{u} \\\psi_d\end{array} } \right)$. Moreover, since $\psi_{u,d}$ are four-component 
spinors, $\psi (x)$ is an eight-component spinor, $\psi^T(x)=(\psi_{u1},...,\psi_{u4},\psi_{d1},...,\psi_{d4})$. (For simplicity, here we ignore the fact that $\psi (x)$ is also a color triplet.) So the path integration measure ${\cal D}\overline\psi 
{\cal D}\psi$ has in detail the following form
\begin{eqnarray}
{\cal D}\overline\psi 
{\cal D}\psi\equiv\prod_{\alpha=1}^8{\cal D}\overline\psi_\alpha\prod_{\beta=1}^8{\cal D}\psi_\beta=\prod_{\alpha=1}^4{\cal D}\overline\psi_{u\alpha}\prod_{\beta=1}^4{\cal D}\psi_{u\beta}\prod_{\alpha=1}^4{\cal D}\overline\psi_{d\alpha}\prod_{\beta=1}^4{\cal D}\psi_{d\beta}.\label{C0}
\end{eqnarray}
First of all, we note that at zero chemical potentials the Lagrangian 
$L_{NJL}$ (\ref{1}) is invariant with respect to the transformations $\psi (x)\to \psi'(x)=
U(\vec\theta)\psi (x)$, where the 2$\times$2 unitary matrix $U(\vec\theta)$ in the two-flavor space 
looks like
\begin{eqnarray}
U(\vec\theta)\equiv U(\theta_1,\theta_2,\theta_3)=\exp (i\vec\tau\cdot\vec\theta)=\cos |\theta|+\frac{i\vec\tau\cdot\vec\theta}{|\theta|}\sin |\theta|=\left (\begin{array}{cc}
\cos |\theta|+\frac{i\theta_3}{|\theta|}\sin |\theta|,& \frac{i\theta_1+\theta_2}{|\theta|}\sin |\theta|\\
\frac{i\theta_1-\theta_2}{|\theta|}\sin |\theta|~,& \cos |\theta|-\frac{i\theta_3}{|\theta|}\sin |\theta|
\end{array}\right ),\label{C1}
\end{eqnarray}
and $\vec\tau\cdot\vec\theta=\tau_1\theta_1+\tau_2\theta_2+\tau_3\theta_3$, $|\theta|^2=\theta_1^2+\theta_2^2+\theta_3^2$. In total, matrices $U(\vec\theta)$ 
form $SU_V(2)$ group. 

It is also well known that $L_{NJL}$ is invariant under (axial) 
transformations $\psi (x)\to \psi'(x)=U_A(\vec\omega)\psi (x)$, where the 2$\times$2 unitary matrix $U_A(\vec\omega)$ in the two-flavor space looks like
\begin{eqnarray}
U_A(\vec\omega)=\exp (i\gamma^5\vec\tau\cdot\vec\omega)=\cos |\omega|+
\frac{i\gamma^5\vec\tau\cdot\vec\omega}{|\omega|}\sin |\omega|,\label{C2}
\end{eqnarray}
and these transformations form the so-called axial $SU_A(2)$ group. 
We have presented expressions for the matrices $U(\vec\theta)$ and $U_A(\vec\omega)$ in a two-dimensional flavor space $\psi=\left( {\begin{array}{c}
\psi_{u} \\\psi_d\end{array} } \right)$. In reality, of course, they act in the space of 8-component Grassmann spinors, $\psi^T(x)=(\psi_{u1},...,\psi_{u4},\psi_{d1},...,\psi_{d4})$, where they look like some 8x8 unitary matrices with determinant equal to one. Now suppose that $\psi (x)\to \psi'(x)=
U(\vec\theta)\psi (x)$, where $U(\vec\theta)\in SU_V(2)$. Then 
we have 
\begin{eqnarray}
{\cal D}\overline\psi' 
{\cal D}\psi'\equiv\prod_{\alpha=1}^8{\cal D}\overline\psi'_\alpha\prod_{\beta=1}^8{\cal D}\psi'_\beta
=\prod_{\alpha=1}^8{\cal D}\overline\psi_\alpha (\det U^\dagger(\vec\theta))^{-1}
\prod_{\beta=1}^8{\cal D}\psi_\beta (\det U(\vec\theta))^{-1}=
{\cal D}\overline\psi {\cal D}\psi,\label{C3}
\end{eqnarray}
i.e. the path integral measure ${\cal D}\overline\psi 
{\cal D}\psi$ (see, e.g., in Eq. (\ref{a8})) is invariant under transformations from 
$SU_V(2)$ group. Here we have used the above mentioned relations $\det U(\vec\theta)U^\dagger(\vec\theta)=1$.
(If we take into account that $\psi (x)$ is also a color triplet, then, using the same method, it is easy to prove that the path integral measure ${\cal D}\overline\psi 
{\cal D}\psi$ is invariant with respect to transformations $\psi (x)\to \psi'(x)=
U\psi (x)$, where $U\in SU_c(3)$.)

In a similar way it is possible to show that this measure is also invariant under arbitrary 
transformation $U_A(\vec\omega)\in SU_A(2)$, where $U_A(\vec\omega)$ is presented in Eq. 
(\ref{C2}). Indeed, suppose that $\psi'(x)=U_A(\vec\omega)\psi (x)$. Then due to an evident relation
$\gamma^0\exp(i\gamma^5\vec\omega\cdot\vec\tau)=\exp(-i\gamma^5 \vec\omega\cdot\vec\tau)\gamma^0$ we have 
$\overline\psi'(x)=
\overline\psi (x)U_A(\vec\omega)$ and hence the following path integration measure transformation
\begin{eqnarray}
{\cal D}\overline\psi' 
{\cal D}\psi'=
\prod_{\alpha=1}^8{\cal D}\overline\psi_\alpha
\prod_{\beta=1}^8{\cal D}\psi_\beta (\det U_A(\vec\omega))^{-2}=\prod_{\alpha=1}^8{\cal D}\overline\psi_\alpha \prod_{\beta=1}^8{\cal D}\psi_\beta \;e^{-iN_c{\rm tr}\{\tau^a\}\,\int d^4 x\omega_a(x)\frac{{\cal A}
}{2}}=
{\cal D}\overline\psi {\cal D}\psi,\label{C4}
\end{eqnarray}
where ${\cal A}=-2{\rm tr}[\gamma_5]\delta(x - x)$ is the same expression as for chiral anomaly of $U_A(1)$ and it is finite if we use regularization. We have also used that there is additional factor ${\rm tr}(\tau^a)=0$ that is equal to zero. Hence  the path integral 
measure ${\cal D}\overline\psi {\cal D}\psi$ is invariant under transformations from $SU_A(2)$ group, in addition. 

Now we can make a very important conclusion that the integration measure ${\cal D}\overline\psi {\cal D}\psi$
is invariant with respect to arbitrary transformations $\psi (x)\to\psi'(x)=U(\vec\theta)U_A(\vec\omega)\psi (x)$.
Supposing here that $\vec\theta=-\vec\omega=(\pi/4,0,0)$, it is easy to conclude that path integration measure 
is invariant with respect to discrete Pauli-Gursey transformation (\ref{12}) when $\psi (x)\to\psi'(x)=
\psi_R(x)+i\tau_1\psi_L(x)$. In addition, it is rather evident that the measure of integration over auxiliary meson fields ${\cal D}\sigma \prod_{i=1}^3{\cal D}\pi_i$ in generating functional (\ref{a3}) is also invariant under the PG transformation (\ref{21}) of these fields.

$\bullet$ Now, based on this result, we are ready to consider the question of how the dual symmetry (\ref{20})-(\ref{21}) of the auxiliary Lagrangian $\widetilde L_{NJL}$ (\ref{3}) is reflected in the properties of the effective action (\ref{aa6}) of the NJL model (\ref{1}) and its thermodynamic potential.

Dual symmetry (\ref{20})-(\ref{21}) of the auxiliary Lagrangian $\widetilde L_{NJL}$ (\ref{3}) leads to the fact that 
both the total effective action $\Gamma (\sigma_{cl},\pi_{a_{cl}})$ (\ref{a6}) and the total 
thermodynamic potential $\Omega (\sigma_{cl},\pi_{a_{cl}})$ (\ref{7}) of the model  are 
symmetric with respect to the transformation 
\begin{eqnarray}
\sigma_{cl}(x)\to-\pi_{1_{cl}}(x),~~\pi_{1_{cl}}(x)\to\sigma_{cl}(x),~~\pi_{2_{cl}}(x)\to\pi_{3_{cl}}(x),~~
\pi_{3_{cl}}(x)\to-\pi_{2_{cl}}(x).
\label{C5}
\end{eqnarray}
In terms of variables $\Sigma$ and $\Pi$ (see the text before Eq. (\ref{8})) it looks like 
$\Sigma\leftrightarrow\Pi$ (and simultaneously $\mu_I\leftrightarrow\mu_{I5}$). 

First of all, we consider the transformation of the functional 
${\cal W}_{{\cal M}}(J_a)$ (it is presented in the expression for the generated functional (\ref{a3})) under the dual transformation (\ref{20})-(\ref{21}). And as a first step, let us swap the chemical potential values, $\mu_I\leftrightarrow\mu_{I5}$, in the chemical potential term $\cal M$  defined by Eq. (\ref{200}). The obtained functional we denote by 
${\cal W}_{\widetilde{\cal M}}(J_a)$, where $\widetilde{\cal M}=\frac{\mu_B}{3}+\frac{\mu_{I5}}2\tau_3+\frac{\mu_{I}}2\gamma^5\tau_3+\mu_5\gamma^5
$. Now, in the functional integral ${\cal W}_{\widetilde{\cal M}}(J_a)$ (it is no more than expression (\ref{a3}), in which  
${\cal M}\to\widetilde{\cal M}$), we change the integration variables, both spinor and auxiliary scalar fields, by the dual transformations (\ref{20})-(\ref{21}).
When performing this procedure, it becomes clear that, since the measure of the functional integral does not change, the original functional ${\cal W}_{\widetilde{\cal M}}(J_a)$ is transformed according to the following rule:
\begin{eqnarray}
{\cal W}_{{\cal M}}(J_a)={\cal W}_{\widetilde{\cal M}}(J'_a),
\label{C6}
\end{eqnarray}
where $J_0'=J_1$, $J_1'=-J_0$, $J_2'=-J_3$, $J_3'=J_2$. According to the left- and right-hand sides 
of Eq. (\ref{C6}), it is possible to introduce, using a definition (\ref{a6}), two formally 
different effective actions,
\begin{eqnarray}
\Gamma_{{\cal M}}(\sigma_{cl},\pi_{a_{cl}})={\cal W}_{{\cal M}}(J_a)
-\int d^4x\left\{\sigma_{cl}(x)J_0(x)+\sum_{a=1}^3
\pi_{a_{cl}}(x)J_a(x)\right\},\label{C71}\\
\Gamma_{\widetilde{\cal M}}(\sigma'_{cl},\pi'_{a_{cl}})={\cal W}_{\widetilde{\cal M}}(J'_a)
-\int d^4x\left\{\sigma'_{cl}(x)J'_0(x)+\sum_{a=1}^3
\pi'_{a_{cl}}(x)J'_a(x)\right\},\label{C7}
\end{eqnarray}
where, by definition (\ref{a6}), we have 
\begin{eqnarray}
\sigma_{cl}(x)=\frac{\delta{\cal W}_{{\cal M}}(J_a)}{\delta J_0(x)},~~
\pi_{b_{cl}}(x)=\frac{\delta{\cal W}_{{\cal M}}(J_a)}{\delta J_b(x)},~~
\sigma'_{cl}(x)=\frac{\delta{\cal W}_{\widetilde{\cal M}}(J'_a)}{\delta J'_0(x)},~~
\pi'_{b_{cl}}(x)=\frac{\delta{\cal W}_{\widetilde{\cal M}}(J'_a)}{\delta J'_b(x)}.
\label{C9}
\end{eqnarray}
But from the last relations and after taking into account (\ref{C6}) one can, for example, obtain
\begin{eqnarray}
\sigma'_{cl}(x)=\frac{\delta{\cal W}_{\widetilde{\cal M}}(J'_a)}{\delta J'_0(x)}=
\frac{\delta{\cal W}_{{\cal M}}(J_a)}{\delta J'_0(x)}=
\int d^4z\frac{\delta{\cal W}_{{\cal M}}(J_a)}{\delta J_b(z)}\frac{\delta J_b(z)}
{\delta J'_0(x)}=\pi_{1_{cl}}(x),
\label{C10}
\end{eqnarray}
where it is also necessary to use the relation $\frac{\delta J_b(z)}{\delta J'_0(x)}=
\delta_{b1}\delta(z-x)$ (the connection between primed sources $J'_b$ and unprimed $J_a$ is given after 
Eq. (\ref{C6})). In a similar way it is possible to find that 
\begin{eqnarray}
\pi'_{1_{cl}}(x)=-\sigma_{cl}(x),~~\pi'_{2_{cl}}(x)=-\pi_{3_{cl}}(x),~~\pi'_{3_{cl}}(x)=
\pi_{2_{cl}}(x),
\label{C11}
\end{eqnarray}
and using these relations one can easy to see that 
\begin{eqnarray}
\sigma_{cl}(x)J_0(x)+\sum_{a=1}^3
\pi_{a_{cl}}(x)J_a(x)=\sigma'_{cl}(x)J'_0(x)+\sum_{a=1}^3
\pi'_{a_{cl}}(x)J'_a(x).\label{C12}
\end{eqnarray}
As a consequence, it follows from Eq. (\ref{C6}) that effective actions (\ref{C71}) and (\ref{C7}) are equal to each other, $
\Gamma_{{\cal M}}(\sigma_{cl},\pi_{a_{cl}})=
\Gamma_{\widetilde{\cal M}}(\sigma'_{cl},\pi'_{a_{cl}})$. 
Of course, the same property is valid for the their TDPs, 
\begin{eqnarray}
\Omega_{{\cal M}}(\sigma_{cl},\pi_{a_{cl}})=
\Omega_{\widetilde{\cal M}}(\sigma'_{cl},\pi'_{a_{cl}}). \label{C13}
\end{eqnarray}
It is shown in Appendix A that due to the $U_{I_3}(1)$ and $U_{AI_3}(1)$ invariance of the NJL model (\ref{1}), the TDP  $\Omega_{{\cal M}}(\sigma_{cl},\pi_{a_{cl}})$ depends on the invarians $\Sigma=\sqrt{\sigma_{cl}^2+\pi_{3_{cl}}^2}$ and $\Pi=\sqrt{\pi_{1_{cl}}^2+\pi_{2_{cl}}^2}$, respectively. But 
the second TDP $\Omega_{\widetilde{\cal M}}(\sigma'_{cl},\pi'_{a_{cl}})$ depends effectively on $\Sigma'=\sqrt{\sigma_{cl}^{\prime 2}+\pi_{3_{cl}}^{\prime 2}}$ and $\Pi'=\sqrt{\pi_{1_{cl}}^{\prime 2}+\pi_{2_{cl}}^{\prime 2}}$. Using relations 
(\ref{C11}) and (\ref{C12}), we see that $\Sigma'=\Pi$, whereas $\Pi'=\Sigma$. So we see from Eq. (\ref{C13}) that full TDP of the NJL model (\ref{1}) satisfies the dual symmetry relation 
\begin{eqnarray}
\Omega_{{\cal M}}(\Sigma,\Pi)=
\Omega_{\widetilde{\cal M}}(\Pi,\Sigma), \label{C14}
\end{eqnarray}
i.e. it is the same as in the mean-field approximation (\ref{9}).
{\vspace{0.2cm}}

Concerning the NJL model (\ref{100}) with a color superconducting channel of quark interaction, we should note first of all that its auxiliary Lagrangian $\widetilde L_{CSC}$ (\ref{CSC}) is invariant with
respect to dual transformations $\cal D$ of both spinor fields (\ref{20}) and auxiliary meson fields (\ref{21}). But at the same time, as it was shown in Section IV, the dual transformation $\cal D$
leaves the diquark scalar fields $\Delta_a(x)$ and $\Delta^*_b(x)$ unchanged (see in Eqs. (\ref{300}) and 
(\ref{310})). Hence, in this case it is easy to show, using the same methods as for considering the dual properties of the simplest NJL model (\ref{1}) (see above in the present Appendix), that the effective action $\Gamma_{CSC}(\sigma_{cl},\pi_{a_{cl}}, \Delta_{k_{cl}},
\Delta^*_{k'_{cl}})$ of this NJL model will be invariant under transformations (\ref{C10}) and (\ref{C11}), 
supplemented by the condition that the classical diquark fields 
$\Delta_{k_{cl}}$ and $\Delta^*_{k'_{cl}}$ remain unchanged. At the same time it is also necessary to permutate the values of chemical potentials $\mu_I$ and $\mu_{I5}$. 

\section{Regularization and dual invariance}
\label{ApD}

The regularization is inevitable in effective models such as NJL model and regularization parameter cannot be removed, hence, the regularization scheme is actually the part of the model definition.
Let us show that regularization procedure does not break the invariance of measure in path integral and hence the dual property of effective model.

One can choose in principle a regularization that would break the invariance of the Lagrangian and hence the duality property of the model would be broken. But it says only that this regularization procedure is not the best to say the least if one is considering the dual properties.
Let us take, for example, 3d momentum cut-off regularization scheme. 

It is instructive to work in the momentum space
$$
\psi(x) = \int \frac{dp^0}{2\pi} \int \frac{d^3p}{(2\pi)^3} \; e^{i p \cdot x} \; \tilde{\psi}(p^0, \mathbf{p})
$$
The measure in path integral with 3-momentum cutoff regularization could be written as
$$
\mathcal{D}\bar{\psi}\,\mathcal{D}\psi = \prod_{|\mathbf{p}|<\Lambda} \prod_{p^0} \prod_{\alpha} d\bar{\tilde{\psi}}_\alpha(p^0,\mathbf{p}) \; d\tilde{\psi}_\alpha(p^0,\mathbf{p})
$$
Let us consider dual transformation (\ref{10}), in the momentum space it would look as follows
$$
PG:~~\tilde{\psi}_R(p^0, \mathbf{p})\to\tilde{\psi}_R(p^0, \mathbf{p}),~~~\tilde{\psi}_L(p^0, \mathbf{p})\to i\tau_1\tilde{\psi}_L(p^0, \mathbf{p}),\;\;\;\;\;\;\;
\tilde{\psi}_R(p^0, \mathbf{p})\equiv\Pi_+\tilde{\psi}(p^0, \mathbf{p}),~~~\tilde{\psi}_L(p^0, \mathbf{p})\equiv\Pi_-\tilde{\psi}(p^0, \mathbf{p}).
$$
It is obvious that though 3-momentum cutoff regularization breaks Lorentz invariance it does not change decompositions into left-handed and right-handed fermions. 

Now for generality let us consider the more general transformation $\psi(x)\to\psi'(x)=U(\vec\theta)U_A(\vec\omega)\psi (x)$ discussed in appendix C at $\vec\theta=-\vec\omega=(\omega,0,0)$ (at $\omega=-\pi/4$ one would get dual transformation).
For small values of parameter $\omega$ it has the following form
\begin{eqnarray}
\psi_R(x)\to\psi_R'(x)=
\psi_R(x),\;\;\;\;\;\psi_L(x)\to\psi_L'(x)=
\psi_L(x)-2i\tau_1\omega\,\psi_L(x)
\end{eqnarray}
In general one has for the dual transformation in momentum space 
$$
\tilde{\psi}_R'(p) = \tilde{\psi}_R(p),\;\;\;\;\;\tilde{\psi}'_L(p) = \tilde{\psi}_L(p) - 2i \int \frac{d^4q}{(2\pi)^4} \, \tilde{\omega}(p-q) \, \tau_1  \, \tilde{\psi}_L(q)
$$
where $\tilde{\omega}(p)=\int d^4x \, e^{-ipx}\omega$.
But if $\omega=const$ as in our case then one has
$\tilde{\omega}(p-q) = \omega \, (2\pi)^4 \, \delta^{(4)}(p-q)$.

And one can get for the transformation
$$
\tilde{\psi}'_L(p) = \tilde{\psi}_L(p) - 2i \omega \, \tau_1  \, \tilde{\psi}_L(p)
$$
One can see that the transformation does not mix various momentum modes of fermion field $\psi$ and if we cutoff the modes at some level $\Lambda$ the measure remains invariant under this transformation and regularization procedure of effective model does not stand in the way.

\end{document}